\documentclass[fleqn,usenatbib]{mnras}

\usepackage{newtxtext,newtxmath}

\usepackage[T1]{fontenc}
\usepackage{ae,aecompl}

\usepackage{graphicx}	
\usepackage{amsmath}	

\title[Ultralow Limit]{A new MWA limit on the 21~cm Power Spectrum at Redshifts $\sim$ 13 $-$ 17}

\author[Yoshiura et al.]{
S.~Yoshiura,$^{1,2,3,4}$\thanks{e-mail: syoshiura@unimelb.edu.au},
B.~Pindor,$^{1,4}$ 
J.L.B.~Line,$^{4,5}$,
N.~Barry$^{1,4}$,
C.~M.~Trott,$^{4,5}$, 
A.~Beardsley$^{6}$,
\newauthor
J.~Bowman$^{7}$,
R.~Byrne$^{8}$,
A.~Chokshi$^{1,4}$,
B.~J.~Hazelton$^{8,9}$,
K.~Hasegawa$^{10}$,
E.~Howard$^{11,12}$,
\newauthor
B.~Greig$^{1,4}$,
D.~Jacobs$^{7}$,
C.~H.~Jordan$^{4,5}$,
R.~Joseph$^{4,5}$,
M.~Kolopanis$^{7}$,
C.~Lynch$^{4,5}$,
\newauthor
B.~McKinley$^{4,5}$,
D.~A.~Mitchell$^{4,11}$,
M.~F.~Morales$^{8}$,
S.~G.~Murray$^{7}$,
J.~C.~Pober$^{13}$,
\newauthor
M.~Rahimi$^{1,4}$,
K.~Takahashi$^{2,14,15}$,
S.~J.~Tingay$^{4,5}$,
R.~B.~Wayth$^{4,5}$,
R.~L.~Webster$^{1,4}$,
\newauthor
M.~Wilensky$^{8}$,
J.~S.~B.~Wyithe$^{1,4}$,
Z.~Zhang$^{13}$,
Q.~Zheng$^{16}$
\\
$^{1}$The University of Melbourne, School of Physics, Parkville, VIC 3010, Australia\\
$^{2}$Kumamoto University, Faculty of Advanced Science and Technology, Kumamoto, Kumamoto 860-8555, Japan\\
$^{3}${Mizusawa VLBI Observatory, National Astronomical Observatory Japan, 2-21-1 Osawa, Mitaka, Tokyo 181-8588, Japan}\\
$^{4}$ARC Centre of Excellence for All Sky Astrophysics in 3 Dimensions (ASTRO-3D)\\
$^{5}$International Centre for Radio Astronomy Research, Curtin University, Perth, WA 6845,Australia\\
$^{6}$Winona State University Department of Physics, Winona, Minnesota 55987, USA\\
$^{7}$Arizona State University, School of Earth and Space Exploration, Tempe, AZ 85287, USA\\
$^{8}$University of Washington, Department of Physics, Seattle, WA 98195, USA\\
$^{9}$University of Washington, eScience Institute, Seattle, WA 98195, USA\\
$^{10}$Division of Particle and Astrophysical Science, Graduate School of Science, Nagoya University, Chikusa, Nagoya 464-8602, Japan\\
$^{11}$CSIRO Astronomy and Space Science (CASS), PO Box 76, Epping, NSW 1710, Australia\\
$^{12}$Macquarie University, Department of Physics and Astronomy, Sydney, 2000, Australia\\
$^{13}$Brown University, Department of Physics, Providence, RI 02912, USA\\
$^{14}$ International Research Organization for Advanced Science and Technology, Kumamoto University, 2-39-1 Kurokami, Kumamoto 860-8555, Japan\\
$^{15}$ National Astronomical Observatory Japan, 2-21-1 Osawa, Mitaka, Tokyo 181-8588, Japan\\
$^{16}$ Shanghai Astronomical Observatory, Chinese Academy of Sciences, 80 Nandan Road, Shanghai 200030, China\\
}
\date{Accepted XXX. Received YYY; in original form ZZZ}

\pubyear{2021}

\begin{document}
\label{firstpage}
\pagerange{\pageref{firstpage}--\pageref{lastpage}}
\maketitle

\begin{abstract}
Observations in the lowest MWA band between $75-100$~MHz have the potential to constrain the distribution of neutral hydrogen in the intergalactic medium at redshift $\sim 13-17$. Using 15 hours of MWA data, we analyse systematics in this band such as radio-frequency interference (RFI), ionospheric and wide field effects. By updating the position of point sources, we mitigate the direction independent calibration error due to ionospheric offsets. Our calibration strategy is optimized for the lowest frequency bands by reducing the number of direction dependent calibrators and taking into account radio sources within a wider field of view. We remove data polluted by systematics based on the RFI occupancy and ionospheric conditions, finally selecting 5.5 hours of the cleanest data. Using these data, we obtain two sigma upper limits on the 21~cm power spectrum in the range of $0.1\lessapprox k \lessapprox 1 ~\rm ~h~Mpc^{-1}$ and at $z$=14.2, 15.2 and 16.5, with the lowest limit being $6.3\times 10^6 ~\rm mK^2$ at $\rm k=0.14 \rm ~h~Mpc^{-1}$ and at $z=15.2$ with a possibility of a few \% of signal loss due to direction independent calibration. 
\end{abstract}




\section{Introduction}

The 21~cm hyperfine transition traces the presence of neutral hydrogen throughout the Universe. In the period between recombination, $z \sim 1100$ and the end of reionisation, $z \sim 6$, the redshifted 21~cm line, observed by the radio telescopes at 1420/(1+$z$)~MHz, encodes both the density and physical state of the intergalactic medium (IGM), as well as the properties and distribution of the earliest luminous sources \citep{Furlanetto2006CosmologyUniverse2}. Thus, detection of the 21~cm line is a powerful probe of the high-$z$ Universe such as the Cosmic Dawn, when the first stars are formed, and the Epoch of Reionisation (EoR), when the UV photons emitted from early galaxies ionised the entire neutral hydrogen in the IGM.

Attempts to detect this signal have included both interferometric attempts to measure fluctuations in the 21~cm brightness temperature during the EoR (MWA \citep{Tingay2013TheFrequencies,Beardsley2016FIRST7,Wayth2018TheOverview2}, LOFAR \citep{VanHaarlem2013LOFAR:Array2}, PAPER \citep{Parsons2010TheResults2}, GMRT \citep{Paciga2013AExperiment2}) as well as global signal experiments which attempt to measure the evolution of the sky-averaged 21~cm brightness temperature (EDGES \citep{Monsalve2017CALIBRATIONREIONIZATION2}, SARAS \citep{Singh2018SARASReionization2}, BIGHORNS \citep{2015PASA...32....4S2}).

 Notable progress in 21~cm line fluctuation analysis has been made at the higher frequency range of $\rm 100<\nu<200~MHz$ in this decade, for example using the data from MWA Phase~I \citep{Dillon2015EmpiricalData,Beardsley2016FIRST7,Barry2019ImprovingObservations2}, MWA Phase~II \citep{2019ApJ...887..141L}, LOFAR \citep{2017ApJ...838...65P}, PAPER \citep{2019ApJ...883..133K}. 
 Latest upper limits on the 21~cm line signal reached $1.8\times 10^3 ~\rm mK^2$ at $z=6.5$ \citep{Trott2020DeepObservations2} and $5.3\times 10^3 ~\rm mK^2$ at $z=9.1$ \citep{ 2020MNRAS.493.1662M}. These limits are starting to constrain the astrophysics of reionisation \citep{2021MNRAS.500.5322G,2021MNRAS.501....1G}. However, the 21~cm power spectrum at the EoR has not been detected due to various systematics. For example, one of the major difficulties is to correct for the Earth's ionosphere \citep{Jordan2017CharacterizationArray2,DeGasperin2018TheObservations2} and which can bias power spectrum estimation \citep{2018ApJ...867...15T}. Radio frequency interference (RFI), which can bias calibration, has to be flagged \citep{Offringa2015TheMitigation,2019MNRAS.483.2207M}. Most importantly, the bright foregrounds are a few orders of magnitude larger than the expected 21~cm signal \citep{2008MNRAS.389.1319J} and become further complicated by interacting with the instrumental response \citep{2019MNRAS.483.2207M}. The power spectrum measurement can also be biased by instrumental systematics such as the beam \citep{2020MNRAS.492.2017J} and  spectral smoothness \citep{2016PASA...33...19T}. Incomplete sky models can introduce calibration errors which can prevent detection of the cosmological power spectrum \citep{2016MNRAS.461.3135B,2019ApJ...875...70B,2020PASA...37...45Z}.

Further interest in 21~cm cosmology was generated by the detection of a broad absorption feature centered at 78~MHz reported by \citet{Bowman2018AnSpectrum}. The unexpected depth ($\sim 500 ~\mathrm{mK}$) and profile of the EDGES detection suggest that fiducial expectations \citep[e.g.][]{Pritchard201221cmCentury2} regarding physical conditions at high redshift may need to be substantially revised \citep{Barkana2018PossibleStars,Fialkov2019SignatureSpectrum2}, although the cosmological origin of this feature remains uncertain \citep{Hills2018ConcernsData2,Singh2019TheSpectrum2}. Interferometric (non)detection of the larger than expected spatial fluctuations in the 21~cm signal which should accompany a large global feature could confirm or constrain the EDGES result. Previous interferometric constraints on the 21~cm fluctuations prior to reionisation have been reported by MWA \citep[hereafter, EW16]{Ewall-Wice2016FirstHeating}, LOFAR \citep{Gehlot2019TheLOFAR}, OVRO-LWA \citep{Eastwood2019TheOVRO-LWA}, and LOFAR-AARTFAAC \citep{2020MNRAS.499.4158G} although none of these results have yet reached the sensitivity required to substantially constrain theoretical models which might predict an EDGES-like global signal. Whether or not the EDGES signal is ultimately confirmed, these investigations emphasise the unique potential of the 21~cm line to probe the $z>15$ Universe and the necessity of preparing for observations with upcoming facilities such as HERA \citep{Deboer2017HydrogenHERA2}, NenuFAR \citep{Zarka2015NenUFAR:Case2}, and SKA1-LOW \citep{Koopmans2014TheArray}.



Unfortunately, all of the same major difficulties which complicate a detection of the EoR signal apply at lower frequencies. Bright radio foregrounds, RFI, ionospheric distortions, instrumental systematics and difficulty of calibration are not only present but invariably more prominent at lower frequencies. In this work, we report on further efforts to measure the 21~cm power spectrum at $z \sim 13-17$ using the Murchison Widefield Array (MWA). We analyse ultralow frequency data, $75<\nu<100$~MHz, using the real time system \citep[RTS;][]{Mitchell2008Real-timeArray} and the Cosmological HI Power Spectrum estimator \citep[CHIPS;][]{Trott2016CHIPS:ESTIMATOR2} for the first time. We mainly investigate the above difficulties and attempt to optimize the calibration setting to mitigate the systematics.

This paper is organised as follows: in Section~\ref{obs_section} we describe the MWA instrument, observations, and our processing pipelines; in Section~\ref{systematics_section} we describe the effect of several important sources of systematic error and mitigation strategies that we have employed to improve our data calibration; and in Section~\ref{results_section} we show power spectrum results using the best data sets; and we conclude in Section~\ref{sec:discussion}.

\section{Observations and data analysis}
\label{obs_section}

\subsection{The MWA}
The MWA \citep{Tingay2013TheFrequencies,Wayth2018TheOverview2} is a low-frequency radio interferometer located at the Murchison Radio Observatory (MRO) in Western Australia . The detection of the EoR is one of the primary science goals of the MWA, along with Solar and Heliospheric radio emission, Galactic and Extragalactic radio astronomy, and time domain studies \citep{Bowman2013ScienceArray,2019PASA...36...50B}.

In this work we consider data from the Phase~I MWA which consisted of 128 tiles distributed semi-randomly over $\sim 3$~km with a densely-concentrated central core. Each tile is a 4 by 4 grid of crossed dipoles 1~m apart on top of a $25~\mathrm{m}^2$ mesh ground plane. The signals from 16 dipoles in each tile are combined in an analog beamformer which can introduce discrete delays for each dipole hence allowing the primary beam of the tile to be electronically steered to different positions on the sky. The MWA receiver system digitises, channelises and selects a 30.72~MHz instantaneous bandwidth \citep{Prabu2015AArray} which is cross-correlated and averaged in the Graphics Processing Unit (GPU) based MWA correlator \citep{Ord2015TheCorrelator}. For these data, the visibilities are recorded at 2~s and 40~kHz resolution.

\subsection{Ultralow observations}

Observations of the MWA EoR target fields in the lowest MWA band were initiated by EW16. We term these observations \textit{ultralow} in contrast to the main EoR low ($138-168$~MHz) and high ($168-197$~MHz) bands. For the observations considered in this work, data were taken in a split band between $75-100$~MHz as well as $152-156$~MHz. Using information from the higher frequency sub-band offers the possibility of improved ionospheric measurements but here we only consider processing of the contiguous $75-100$~MHz data. The EoR target fields are chosen to be cold regions of the radio sky and are designated EoR0 (centered at 0~h RA, -27~deg. DEC) and EoR1 (~4~hr, -30~deg.). The discrete nature of the beamformer delays means that the MWA does not track the target field but rather updates the pointing centre approximately each 30 minutes and allows the target field to drift through. Each such drift scan is called a \textit{pointing}. We also restrict our analysis to the central three pointings and where noted focus on the zenith pointing to reduce variations due to different primary beam shapes. Observations used in this work are summarized in Table~\ref{tab:data_table}.

\begin{table}
	\centering
	\caption{MWA Ultralow observations including zenith and $\pm$ 1 off zenith observations. The observation performed from 2014-09-22 to 2014-10-25. }
	\label{tab:data_table}
	\begin{tabular}{lcr} 
		\hline
		\hline
		 Total data volume EoR0 & 7.4 & hours\\ 
		 Total data volume EoR1 & 6.4 & hours\\ 
		Phase Centre EoR0 & $0^h$ $-27^{\circ}$& \\
		Phase Centre EoR1 & $4^h$ $-30^{\circ}$& \\
		Frequency range& 74.88 $< \nu < $ 100.48 &~MHz\\ 
		 Coarse band channel & 1.28 &~MHz\\ 
		 Frequency resolution & 40 & kHz\\ 
		 Time per snapshot & 112 & s\\ 
		 Time resolution & 2 & s\\ 
		 MWA configuration & Phase~I &\\ 
		\hline
	\end{tabular}
\end{table}



\begin{figure*}
	\includegraphics[width=8cm]{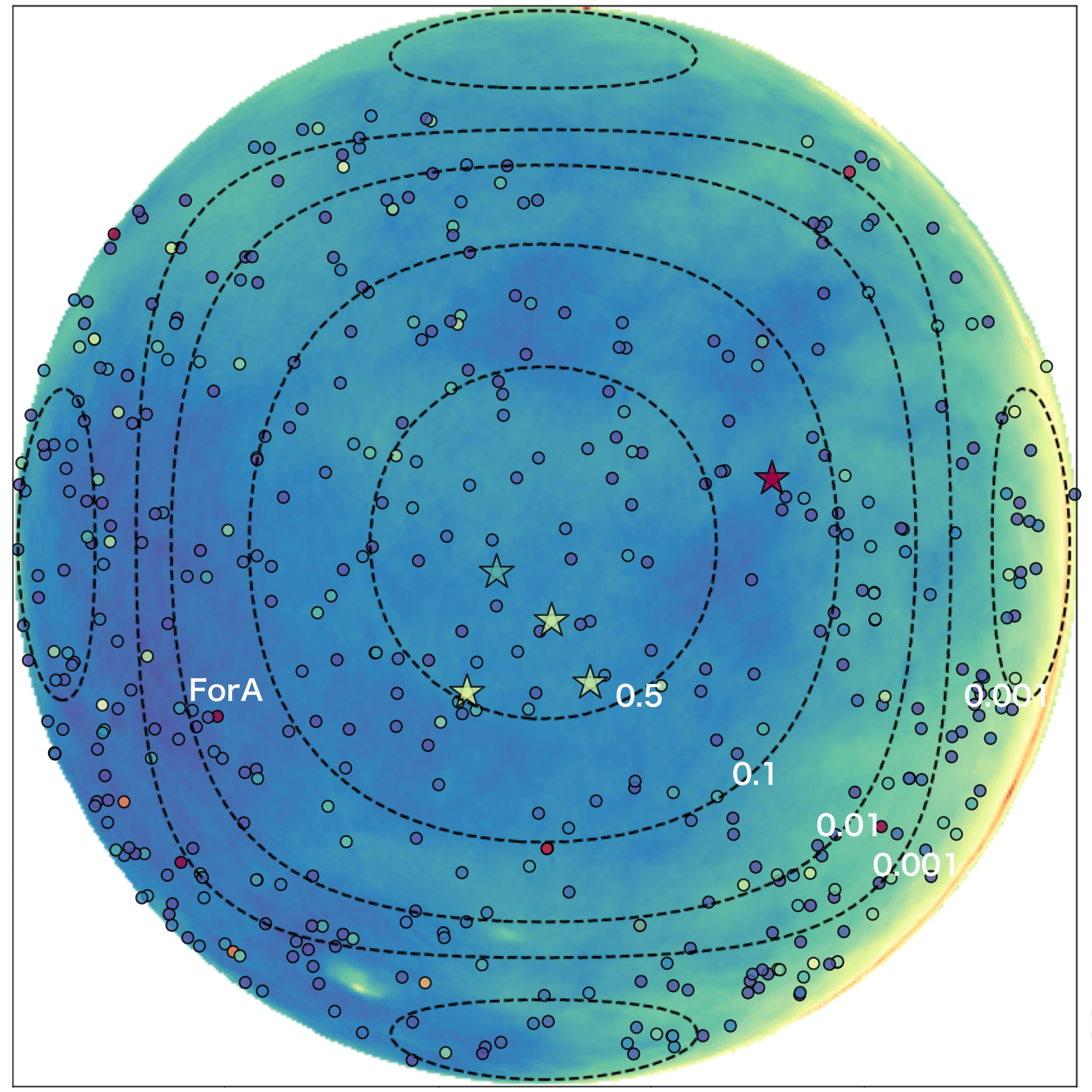}
    \includegraphics[width=8cm]{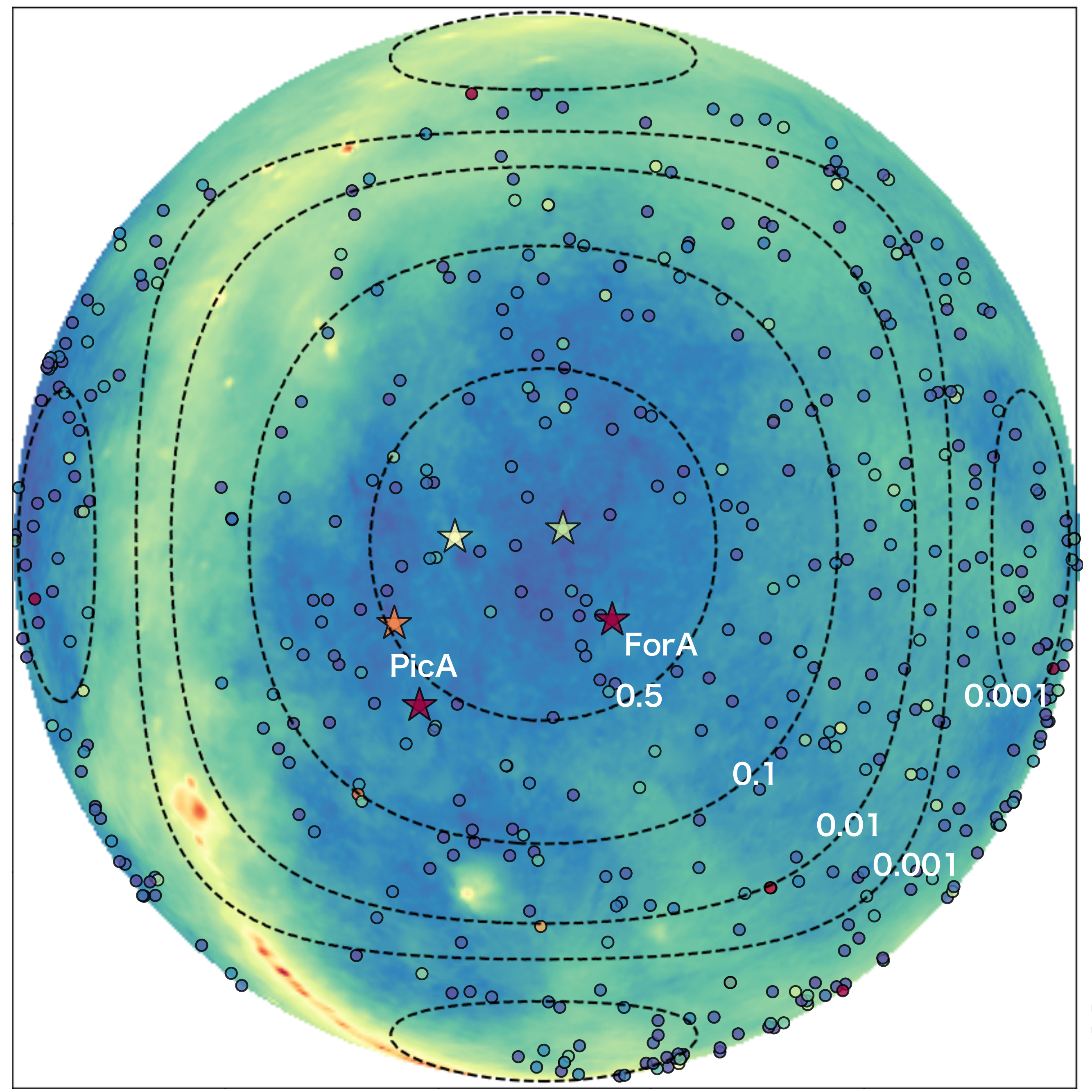}
    \caption{Sky map of EoR0 and EoR1. The EoR0 and EoR1 is centered at (RA, Dec = 0h, -27deg) and (RA, Dec = 4h, -30deg). The contours are MWA's beam response at 87~MHz. As an example, coloured circles indicate the radio sources brighter than 10~Jy at 87~MHz and listed in our calibration source list. The redder (bluer) color indicates bright (faint) source.  The background diffuse emission is improved Haslam map at 408~MHz. The five star marks indicate examples of direction dependent calibrators for zenith observation.  }
    \label{fig:skyimage}
\end{figure*}

Fig.~\ref{fig:skyimage} shows our sky models for the EoR0 and EoR1 fields. The background diffuse emission is the improved Haslam map at 408~MHz \footnote{\url{https://lambda.gsfc.nasa.gov/product/foreground/fg_2014_haslam_408_info.cfm}} \citep{Remazeilles2015AnMap2}, and also we show radio sources brighter than 10~Jy at 87~MHz as example, and the beam response of MWA at 87~MHz.  
Due to the very wide field of view of the MWA at the ultralow frequencies, the bright `A-team' sources Fornax~A and Pictor~A are present in both the EoR0 ane EoR1 observations. In both sets of observations these sources are far from the phase centre, however in EoR1 they are in the main lobe of the primary beam. While outside of the primary beam in EoR0, Fornax~A is still the brightest source above the horizon and Pictor~A is the 680th brightest source at 81~MHz, at 1~apparent~Jy and close to horizon. Note that removing Pictor~A from the visibilities does not improve our power spectrum results for a 10~min integration for EoR0.


\subsection{RTS and CHIPS}

\subsubsection{Calibration}
We processed data through an analysis pipeline designed to measure the 21~cm power spectrum from MWA EoR data. Raw visibilities from the MWA archive are processed into calibrated residual visibilities though the MWA Real-Time System \citep[RTS;][]{Mitchell2008Real-timeArray}. We here describe briefly the RTS calibration procedure. Pre-processing is performed by using COTTER which performs the RFI flagging using AOFLAGGER package \citep{Offringa2015TheMitigation}. The data are recorded as \textit{observations} which are each  112~s of data with 2~s time resolution and 40~kHz frequency resolution. Each such observation is independently processed through the RTS.  

The RTS, a GPU based calibration software, applies a sky-based calibration procedure. The sky model is generated from an all sky catalogue which is created by combining multiple observations using PUMA \citep{Line2017PUMA:Algorithm2}. The base source catalogue includes GLEAM \citep{Wayth2015GLEAM:Survey2}, TGSS images \citep{2017PASA...34...33P}, and MWA extended baseline data to model bright complex sources using shapelets \citep{Line2020ModellingStudy2}. The sources are ranked by beam weighted flux density, where the beam weights are evaluated for all sources at central frequency of entire bandwidth. In this work, the 1000-2000 highest ranked sources within a certain radius from the pointing centre are selected. For the highband, radius of 20 degrees is chosen. However, for ultralow frequency, the beam response is considerably wider, and we empirically set a radius of 60 degrees.

The initial calibration is a direction independent (DI) calibration which combines the 1000-2000 selected sources into a single compound calibrator and solves for each complex antenna gains (Jones matrix). These solutions are performed independently for each coarse channel in parallel and are stored for later processing. We exclude baselines shorter than 20$\lambda$ and taper baselines shorter than 40 $\lambda$ because they are dominated by Galactic diffuse emission which is not modeled in our catalogue. 


Once the DI solution is obtained, the RTS performs direction dependent calibration. The procedure is called peeling and operated within a calibration measurement loop. In order to obtain the Jones matrix and/or to correct for ionospheric effects, the RTS reads a peeling source list which also includes 1000-2000 sources. Using the peeling source list and DI calibration result, all modelled sources are subtracted from the observed visibilities. The model subtracted visibilities are phase shifted to the position of the desired calibrator. Then, the subtracted source model of the calibrator is added to the subtracted visibilities, and the RTS solves the Jones matrix for the direction dependent calibrator for each coarse band. Typically, the five brightest sources are selected as direction dependent calibrators, however, based on the results of section \ref{sec:ddcal}, we used zero DD calibrators for the processing presented in our final limits. For the rest of the calibrator sources, instead of full Jones matrix, a correction of ionospheric displacement and an amplitude scaling are obtained using the entire frequency band. We list our fiducial RTS setting in Table.~\ref{tab:rtsfid}.

\begin{table}
	\centering
	\caption{Our fiducial RTS parameters used for investigating data quality. A direction dependent calibration is performed towards the number of sources listed as DD calibration. Ionospheric phase correction is performed towards the number of sources listed as ionosphere correction and this is also the number of sources to be subtracted. Baselines less than the minimum baseline length are excluded from the calibration. We down weight baselines shorter than the tapered baseline length in the antenna gain calibration. }
	\label{tab:rtsfid}
	\begin{tabular}{ll} 
		\hline
		\hline
		 Number of sources for DD calibration & 5\\ 
		 Number of sources for ionosphere correction & 1000\\ 
		 Amplitude correction & on\\ 
		 Number of sources to be subtracted & 1000\\ 
		 Beam model & Analytic\\ 
		 Minimum baseline length & 20 $\lambda$\\ 
		 Tapered baseline length & 40 $\lambda$\\ 
		\hline
	\end{tabular}
\end{table}

The phase error for the visibility observed by a baseline $u$ due to ionosphere is described as $\exp\{-i \alpha u\lambda^2 \}$, where $\alpha \lambda^2$ is the offset in a direction and  $\lambda$ is observed wavelength. During the peeling step, the ionospheric offset is measured in the direction of each source in the linear regime that $\alpha u\lambda^2 \ll 1$, and is corrected for before source subtraction. This peeling-based ionospheric calibration has been tested at higher frequencies  \citep[][hereafter,  CJ17]{Intema2009IonosphericResults2,Jordan2017CharacterizationArray2}. However, at ultralow frequencies, the linear assumption might be not valid for longer baselines. For a 3~km baseline, the assumption might bias the correction for ionosphere distortion of $\alpha\lambda^2>1'$. The typical ionospheric median offset is approximately 0.15' at 200~MHz for quiet ionosphere data (CJ17). The median value corresponds to approximately 0.3$'$ at 85~MHz, and thus the ionosphere correction should work well for a quiet ionosphere. 


Using the the full Jones matrix or ionospheric calibration, the calibrator is subtracted again, and the RTS continues the peeling for the next source. This direction dependent (DD) calibration is operated with 8~s time resolution. As for the DI calibration, we exclude and taper short baselines to avoid signal loss due to unmodelled diffuse emission. Once the DD calibration is completed for all calibrators and all time steps, the RTS outputs source subtracted residual visibilities which are used for power spectrum estimation as described below.

\subsubsection{Power Spectrum Estimation}
The cylindrical power spectrum (2D power spectrum) is described in 2D Fourier space ($\rm k_{ \parallel}$, $\rm k_{ \perp}$), where $\rm k_{ \perp}$ corresponds to angular scales perpendicular to the line of sight and $\rm k_{\parallel}$ to those the line of sight. Since the foreground has smooth spectrum, the foreground contamination is effectively reduced at higher $\rm k_{\parallel}$ modes while mode mixing caused from interferometer's chromaticity propagates foreground containation to higher $\rm k_{\parallel}$ called foreground (FG) wedge. In the rest of scale at high $\rm k_{\parallel}$ and low $\rm k_{\perp}$, the 21~cm signal has possibility to dominate the foreground contamination and the scales are called the EoR window \cite{Datta2010BrightMeasurements2, 2012ApJ...752..137M, 2012ApJ...757..101T}. The spherically averaged power spectrum (1D power spectrum) at k is evaluated by averaging the 2D power spectrum within the corresponding k-bins.

The power spectrum is calculated by using the Cosmological HI Power Spectrum estimator \citep[CHIPS;][hereafter,  CT16]{Trott2016CHIPS:ESTIMATOR2}. CHIPS has been used in various previous works to analyse MWA data at higher frequencies \cite[e.g.][hereafter, CT20]{Trott2020DeepObservations2}. We mention some recent modifications here. The inverse co-variance weighting method is replaced by an inverse variance estimator as explained in CT20. This implementation is introduced to avoid signal loss due to mis-modelling of foregrounds. Frequency modes were transformed by regular Fast Fourier Transformation (FFT). Although a Least Squares Spectral Analysis (LSSA) was suggested for the frequency transform in CT16, more recently CHIPS omits LSSA to avoid signal loss. 

Due to aliasing from the polyphase filterbank used in the signal processing chain of MWA, 40~kHz and 80~kHz fine channels at centre and edges, respectively, of each 1.28~MHz coarse bands are always flagged. Thus, the spectral behaviour of the output visibilities has a comb structure, and this results in foreground contamination seen as spectral harmonics in the EoR window. To mitigate this effect, a Gaussian processing based kriging method was introduced in CHIPS (CT20). However, we do not use kriging in this work because the kriging does not show clear improvements on the power spectrum at large scales in CT20. 

\section{Systematic Errors and Mitigation}
\label{systematics_section}

In order to assess their quality, we examined several variable aspects of our data. Three main sources of systematic error, namely RFI, the ionosphere and the wide field of view, were evaluated. Here we describe the data quality thresholds implemented based on these evaluations, and the modifications to the standard MWA EoR calibration strategy that were used to improve our power spectrum results. 

\subsection{RFI}

Although the MRO is an extremely radio-quiet site, RFI is unavoidable in the ultralow data, particularly in the FM band (87-108~MHz) where bright terrestrial signals can come over the horizon, be ducted through the troposphere, or be reflected off aircraft or even the Moon \citep{2013AJ....145...23M2}. We investigated the RFI environment in the ultralow band by considering the flagging statistics of AOFlagger as well as SSINS \citep{2019PASP..131k4507W2}, a different RFI detection algorithm developed for EoR data. 

\subsubsection{AOFlagger Occupancy}

\begin{figure}
    \centering
    \includegraphics[width=\columnwidth]{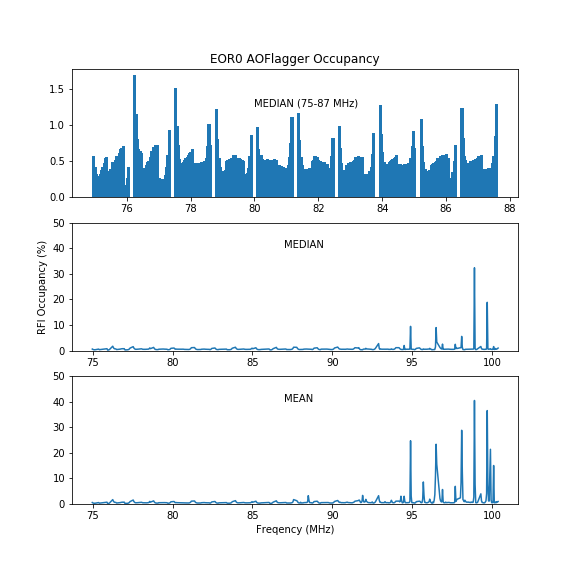}
    \caption{Percentage of samples flagged by AOFlagger as a function of observational frequency for EoR0. Bottom panel shows median occupancy, middle panel shows mean occupancy and top panel focuses on median occupancy below 89~MHz.}
    \label{fig:AOFlagged_Spectrum_EOR0}
\end{figure}

\begin{figure}
    \centering
    \includegraphics[width=\columnwidth]{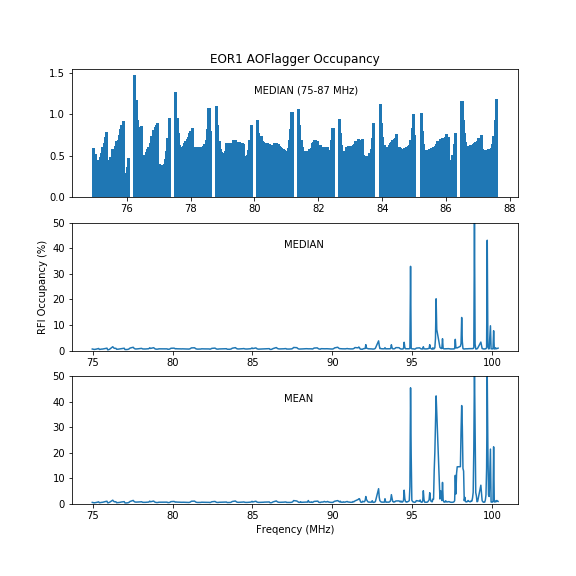}
    \caption{Percentage of samples flagged by AOFlagger as a function of observational frequency for EoR1. Bottom panel shows median occupancy, middle panel shows mean occupancy and top panel focuses on median occupancy below 89~MHz.}
    \label{fig:AOFlagged_Spectrum_EOR1}
\end{figure}

AOFlagger returns a boolean flagging value for every time, frequency and baseline sample to indicate the suspected presence of RFI. For each MWA observation we average these flags in time to determine the flagging occupancy as a function of observing frequency. Figures~\ref{fig:AOFlagged_Spectrum_EOR0} and \ref{fig:AOFlagged_Spectrum_EOR1} show the mean and median occupancy of the 2014 EoR0 and EoR1 data, respectively. Unsurprisingly, there is frequent contamination in the FM band with many channels showing persistent contamination above 10\%. Below the FM band, the occupancy is typically at the $\sim 0.5$ per cent level and clearly shows evidence of the coarse band shape modulating the rate of flagging false positives. Spectral structure introduced by such systematic flagging effects may eventually limit 21~cm observations \citep{Offringa2019TheAnalyses2}. Also notable is the clearly higher ensemble occupancy of EoR1 with respect to EoR0; possibly owing to a change in the underlying visibility statistics caused by the brightness of Fornax~A. In future, fine-tuning of the flagging algorithm is likely warranted but here we note that trials of ignoring the AOFlagger flags below the FM band did not produce improved results indicating that these effects are not limiting our current analysis.    

\begin{figure}
    \centering
    \includegraphics[width=\columnwidth]{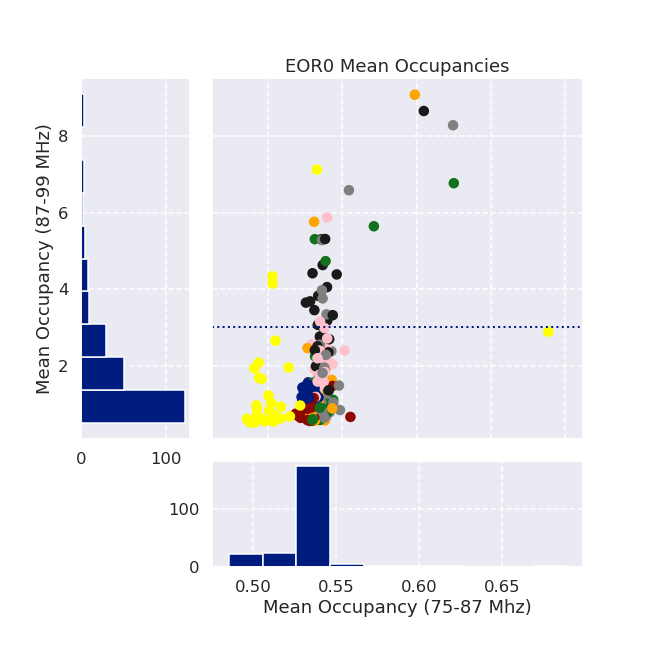}
    \caption{Percentage of samples flagged by AOFlagger for each EoR0 observation. Dotted lines indicate the selection criteria described in the text. All EoR0 observations pass the cut on RFI below the FM band. Colours represent observations from the same night.}
    \label{fig:EOR0_Mean_Occupancies}
\end{figure}

\begin{figure}
    \centering
    \includegraphics[width=\columnwidth]{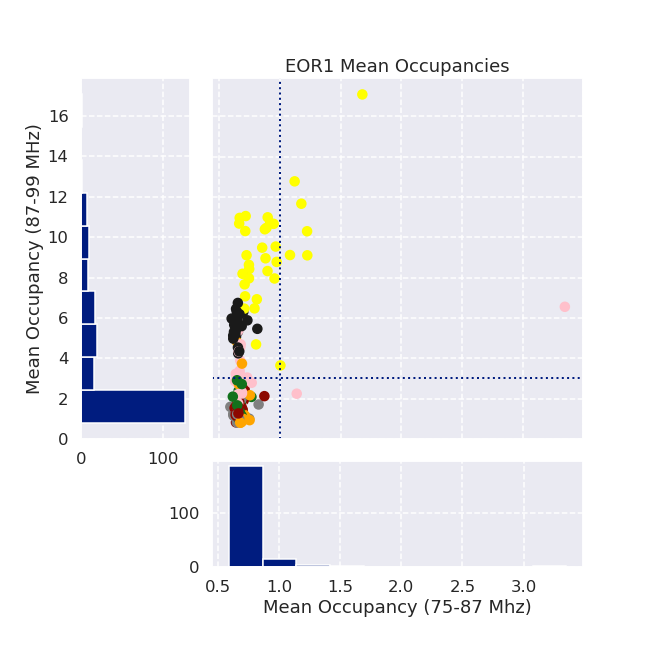}
    \caption{Percentage of samples flagged by AOFlagger for each EoR1 observation. The colours in this do not correspond to the same night of observation as for EoR0.}
    \label{fig:EOR1_Mean_Occupancies}
\end{figure}

We further considered the flagging occupancy of individual observations so as to select data with low RFI for power spectrum measurement. Figures~\ref{fig:EOR0_Mean_Occupancies} and \ref{fig:EOR1_Mean_Occupancies} show the mean AOFlagger occupancy averaged over the top and bottom half of our band for each observation of EoR0 and EoR1 respectively. There is no obvious correlation between the RFI below (75-87~MHz) and in the FM band (87-99~MHz). Based on these distributions we selected low RFI data as having occupancy < 3\% in the FM band and < 1\% below. 

\subsubsection{SSINS Occupancy}

\begin{figure}
    \centering
    \includegraphics[width=\columnwidth]{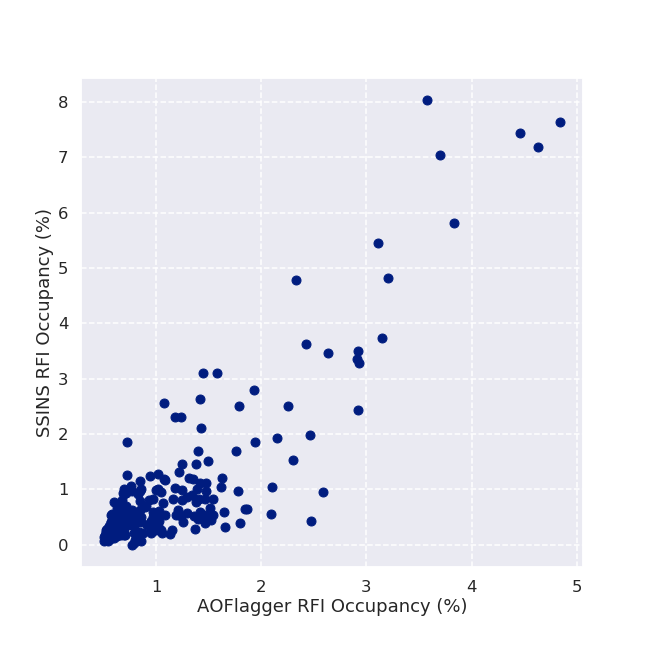}
    \caption{Percentage of samples flagged by AOFlagger vs flagged by SSINS.}
    \label{fig:SSINS_Occupancies}
\end{figure}

We also searched for RFI using SSINS. Unlike AOFlagger which operates on individual baselines, SSINS considers the statistics of an incoherent sum over baselines as a function of time and frequency. In the EoR high band (168 - 198~MHz), SSINS has proven to be particularly sensitive to faint broadband intererence such as digitial television (DTV). \citet{Barry2019ImprovingObservations2} found that using a SSINS-based matched filter to excise observations contaminated by DTV led to a significant improvement in the measured power spectrum limit. Although there is no known equivalent of DTV in the ultralow band, we applied SSINS to our data to search for unknown unknowns. Figure~\ref{fig:SSINS_Occupancies} compares the RFI occupancy as measured by SSINS and AOFlagger. The occupancies are highly correlated and there is no evidence of SSINS-detected RFI which has been missed by AOFlagger. Consequently, we did not add any additional SSINS-based selection to our data.  

\subsection{Ionosphere}

\subsubsection{Ionospheric Conditions}

Earth's ionosphere is a partially ionised gas at altitude of 50~km to 1000~km. The ionosphere refracts incoming radio waves and shift their phase \citep{2017isra.book.....T2}. As a consequence of ionospheric refraction, point source subtraction can leave large residuals since the apparent positions of extra-galactic radio sources are shifted from their true position stored in the source catalogue. Since the positional offset due to the ionosphere is proportional to $\nu^{-2}$, the ionospheric conditions becomes critical at ultralow frequency. \citet{DeGasperin2018TheObservations2} investigated the effect of the ionosphere at 60~MHz. At the low frequencies, 2nd order terms caused by Faraday rotation become non-negligible because the term is proportional to $\nu^{-3}$. The error due to Faraday rotation is less than a few per cent compared to the first order. Thus, we ignore the Faraday rotation effect since our frequencies of interest are higher than 75~MHz, but introducing a method to correct these effects might be required to achieve the detection of 21~cm line at lower frequencies.  


Using the RTS-measured ionospheric offset at each source direction, we can explore the ionospheric conditions for each 2~min snapshot. CJ17 developed  the \textbf{cthulhu}\footnote{\url{https://gitlab.com/chjordan/cthulhu}} package for investigating the ionosphere.\textbf{Cthulhu} defines a metric of ionospheric activity based on the median offset and the anisotropy of ionospheric structure, where the anisotropy is quantified as eigenvalue evaluated using principal component analysis (PCA) to the source offsets. This metric has already been used in previous high-band data analysis (e.g. CT20). 

\begin{figure}
	\includegraphics[width=\columnwidth]{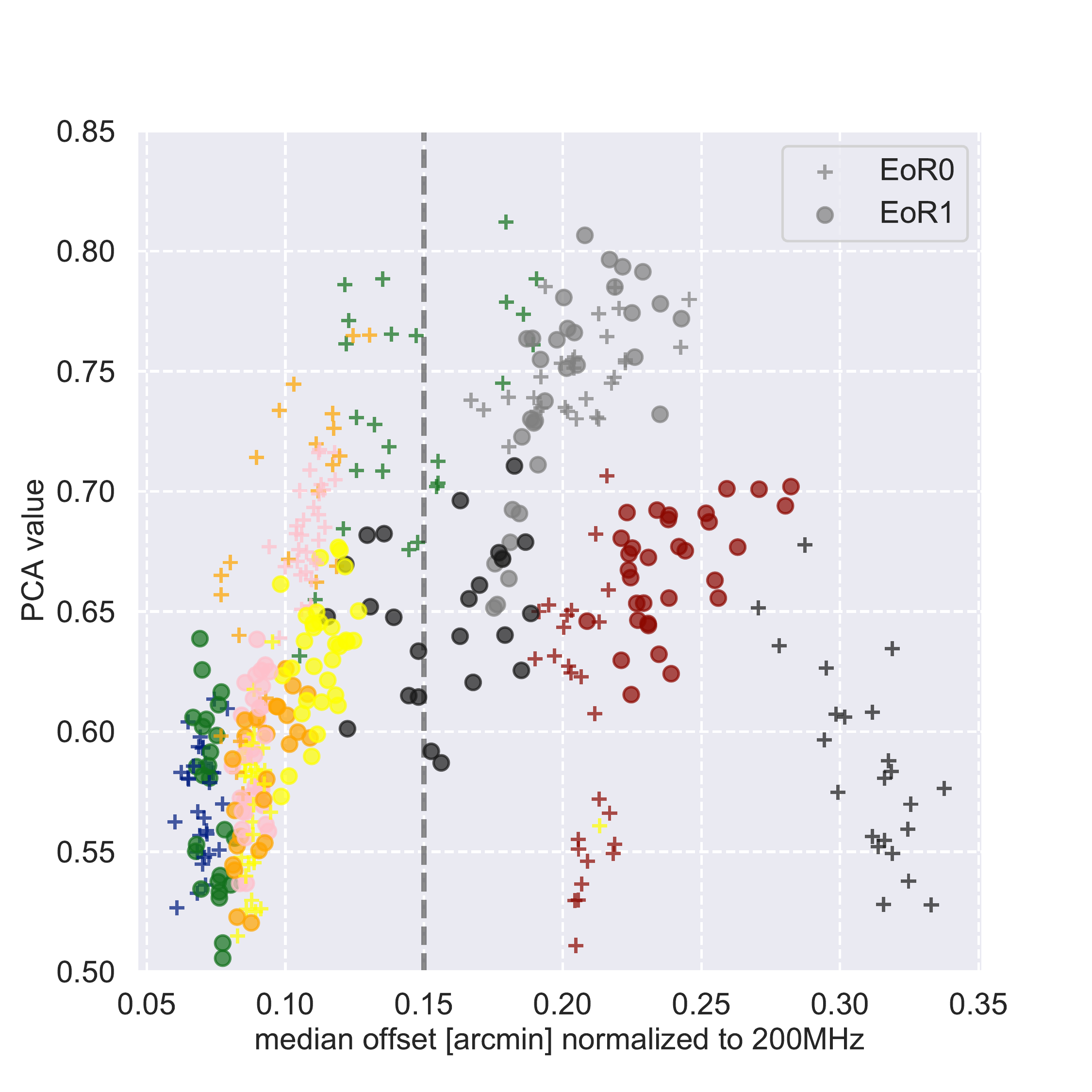}
    \caption{Ionosphere condition at ultralow frequency. The PCA value and the median offset are calculated by using \textbf{cthulhu} from log files of RTS with fiducial setting. Colour indicates different dates of the observation. We set the median offset of 0.15~arcmin as the threshold of ionosphere activity, indicating with vertical dashed line. }
    \label{fig:IONOcondition}
\end{figure}

Fig.~\ref{fig:IONOcondition} shows ionospheric conditions at ultralow frequency. In CJ17, data within the range of median $< 0.15 ~\rm arcmin$ and PCA $<70$ are classified as ionospherically quiet. Note that the median offset is calculated from the measured offsets in the direction of 1000 sources at 75~MHz and the median value is normalized to 200~MHz with $({\rm 75~MHz}/{\rm 200~MHz})^2$. For the EoR1 field, 61\% of data are ionospherically quiet and the figure is 60\% for the EoR0 field. We note, however, that there is no quantitative threshold guaranteed not to contaminate the power spectrum for the ultralow data.

\begin{figure}
\centering
	\includegraphics[width=\columnwidth]{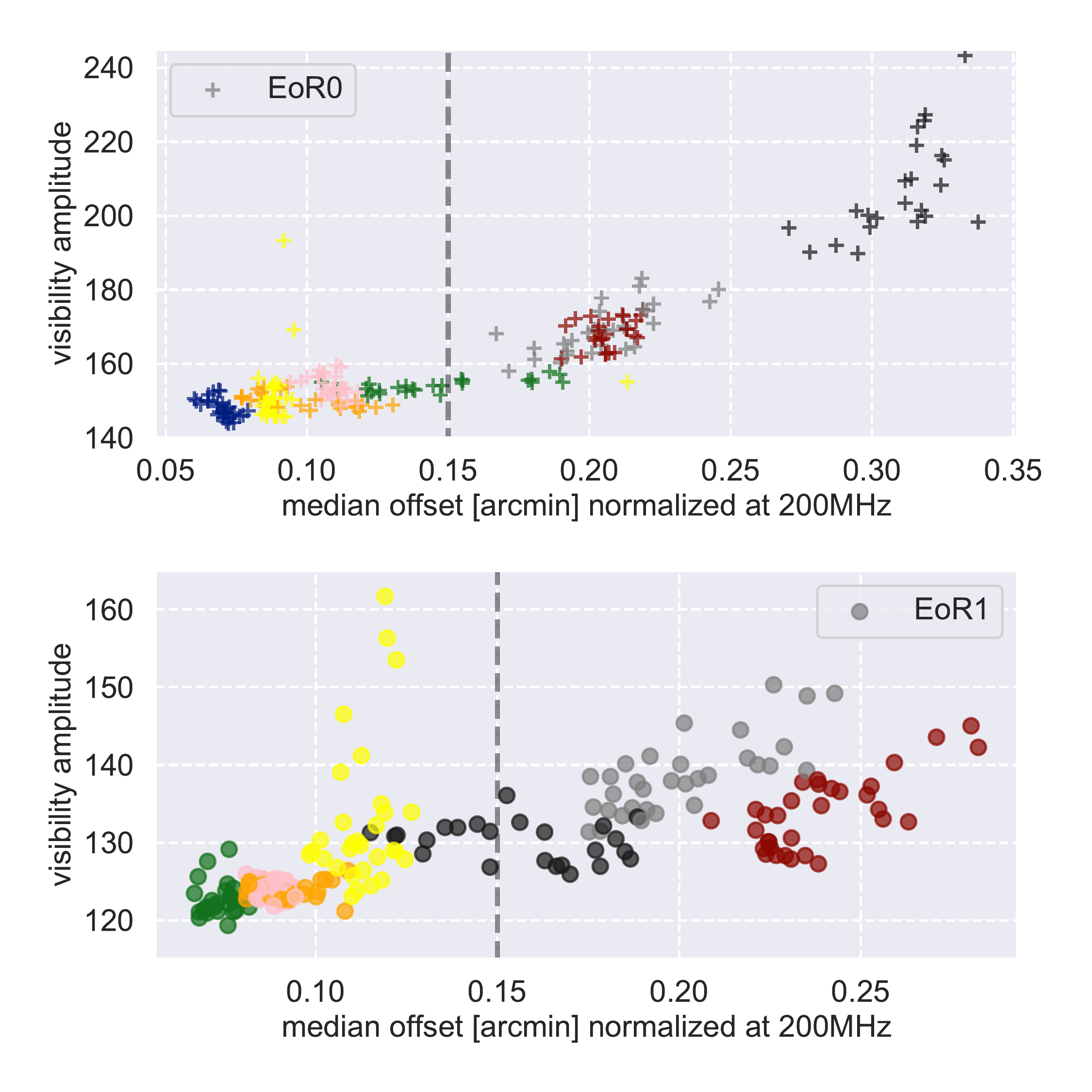}
    \caption{
    The visibility amplitude vs the ionosphere median offset. Different colours indicate different dates of observation. Top and bottom panels show the result of EoR0 and EoR1. The offset correlates with visibility amplitude when the offse is larger than 0.15~arcmin indicated as vertical dashed line. }
    \label{fig:RMSIONO}
\end{figure}

Strong ionospheric contamination can degrade the RTS calibration results. For example, as shown in Fig.~\ref{fig:RMSIONO}, the amplitude of calibrated visibilities  correlates with the median source offset, where the visibility amplitude is calculated as an average over all baselines. Ideally, the visibility amplitudes should not depend on ionospheric conditions. The opposite trend is found in the averaged amplitude of the DI calibration complex gains which are calculated using all tiles and all channels. These results indicate the ionospheric error biases the gains due to effective errors of the patch source model. This figure motivates removing data with median offsets of $>$0.15~arcmin.

\subsubsection{Updating the calibration model}\label{sec:ionoupdate}

Fig.~\ref{fig:RMSIONO} indicates that even for the direction independent calibration (patch calibration), correction for the ionosphere is essential. In order to improve the calibration, we update the patch catalogue based on measured ionospheric offsets. We first measure the ionosphere offsets at each source using full band data via the peeling calibration with 64~s integration time. Next, we shift the position of each source based on the measured offset at 88~MHz. Using the updated source model, we repeat the usual patch (DI) calibration with the RTS for each obsid. 

\begin{figure}
	\includegraphics[width=\columnwidth]{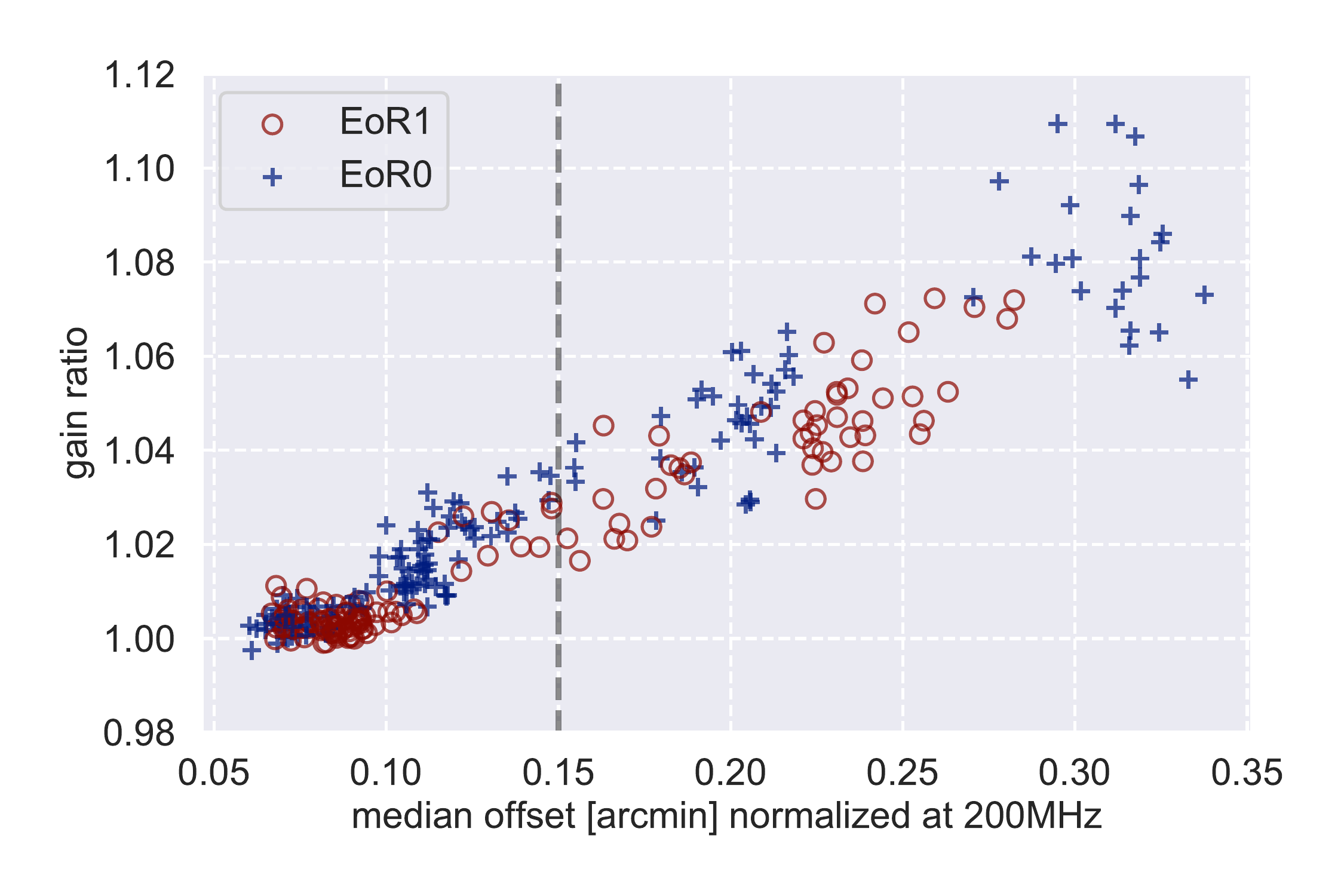}
    \caption{Ratio of gain amplitude with ionosphere patch calibration update and the gain without the update. We calculate median gain amplitude among all tiles at 75~MHz. The ratio shows clear correlation with median ionosphere offset. These are all zenith observation. Threshold of the ionosphere median offset is indicated as vertical dashed line.}
    \label{fig:gainionoupdate}
\end{figure}

Fig.~\ref{fig:gainionoupdate} shows the ratio between the median value of the DI gain result with ionosphere patch calibration and without the update. While the ratio is consistent with one for the data with quiet ionosphere, the gain tends to be underestimated for data with active ionosphere. This result indicates that due to the offset of the apparent position of the brightest sources, the patch catalogue cannot model the sky correctly, and the error of the calibration model results in biased gains. 

\begin{figure}
	\includegraphics[width=\columnwidth]{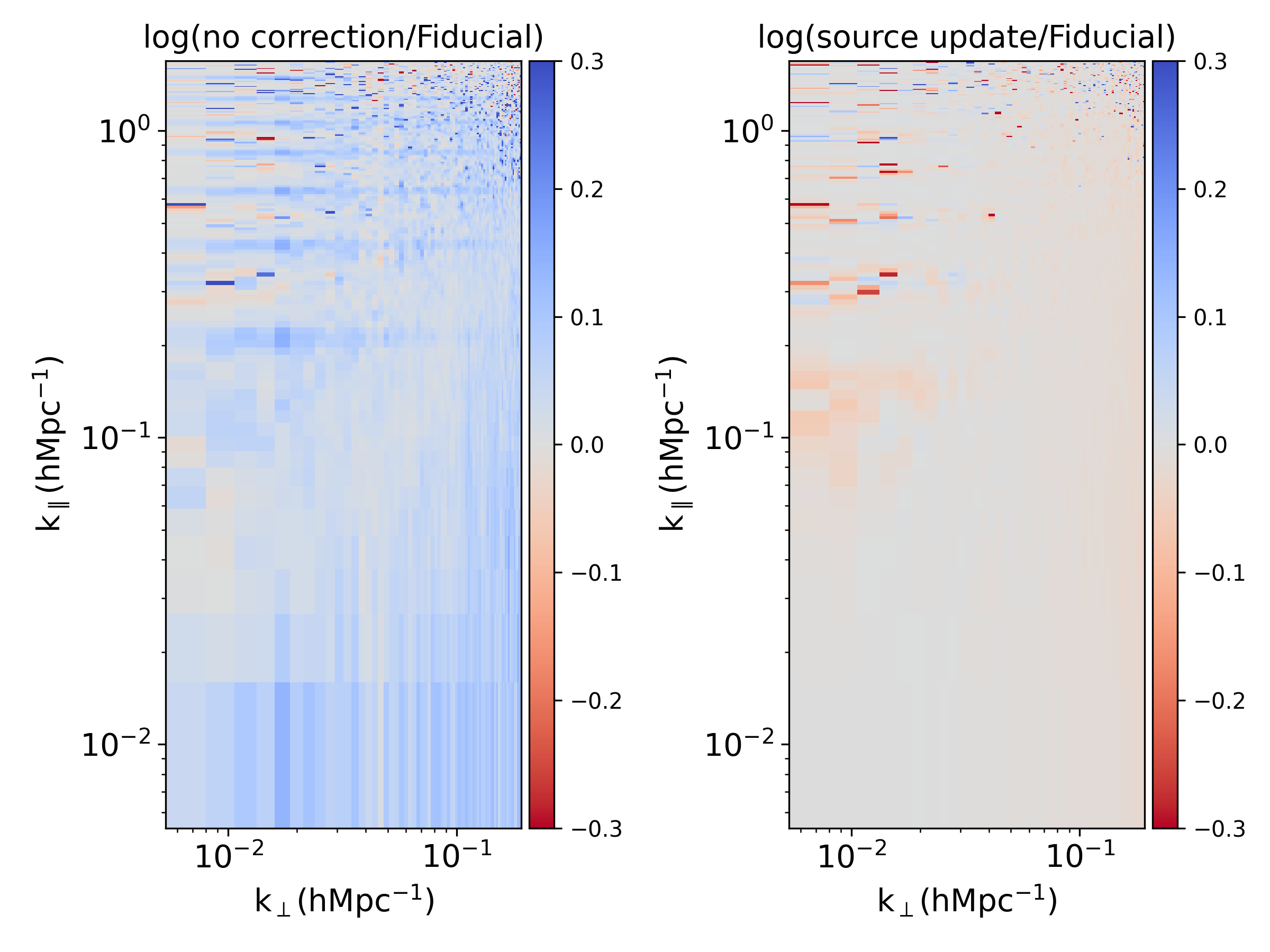}
    \caption{Comparion of power spectrum of different ionosphere correction strategies for the EoR0 field. Left panel shows the logarithm of the ratio of no ionosphere correction to fiducial run. Blue indicates power spectrum without ionosphere correction has larger power than fiducial. Right panel shows the logarithm of the ratio of ionospheric updated source catalogue to fiducial. Red indicates power spectrum with updated source is less than fiducual.  }
    \label{fig:gainionoupdatePS}
\end{figure}

The 2D power spectrum is useful to evaluate the effect of ionosphere correction in 2D Fourier space. In the power spectrum regime, we compare three strategies of ionospheric calibration : (i) fiducial RTS ionosphere correction with original source list, (ii) without ionosphere correction, (iii) RTS with updating the source list. In Fig.~\ref{fig:gainionoupdatePS}, we plot the logarithm of ratio of each power spectrum. Compared to fiducial correction, peeling without ionosphere correction leaves residuals at all scales as shown in left panel of Fig.~\ref{fig:gainionoupdatePS}. The right panel of Fig.~\ref{fig:gainionoupdatePS} shows that the updating source list method slightly improves the power spectrum in the EoR window even in data with good ionospheric conditions. This result indicates that the error of DI calibration results in unwanted frequency structure in gain which propagates the foreground power into the EoR window. The effect is more prominent for data with active ionosphere, but we omit such data from deep integration to reduce systematic errors as similar to  previous works (e.g. CT20). 

We mention that introducing ionospheric offsets in DI calibration is a new procedure in this work. While we show the method can improve the calibration at ultralow frequency, this strategy might be not effective at high band. This is because the improvements are not prominent for quiet data even at the ultralow frequency and ionosphere offset at the high bands will be 4 times lower than the offset at ultralow frequency. This work focus on ultralow frequency, and we leave more quantitative discussion at higher band for future works.

\subsection{Wide Field of View}

The primary beam of the radio telescope becomes wider with $\nu^{-2}$, and the MWA's beam is tremendously large at ultralow frequencies. For example, the beam FWHM is 42 degrees at 87~MHz. Thus, the wide field effects become even more prominent than the EoR high band. In this section, we present analyses related to the beam model, direction dependent calibration, contamination from Galactic plane and point source subtraction.  

\subsubsection{Beam model}

In the fiducial run, we use an analytic MWA beam model. While the analytic model is fairly accurate, the true beam pattern does differ from the analytic model mainly due to mutual coupling effects \citep{Ung2020}. As the MWA beam model is not well tested in the ultralow band, we performed tests comparing an analystical implementation of the beam model in the RTS (which is much faster), to the Fully Embedded Element (FEE) model \citep{Sutinjo2015UnderstandingPolarimetry}, which has been shown to be more accurate at higher frequencies (\citet{2017PASA...34...62S2,2018PASA...35...45L2,2021MNRAS.tmp..200C2}).


The beam model is used to create the calibration model. As the result of the DI calibration, the averaged gain amplitude of the analytic beam model is roughly 1\% larger than that of the FEE model. We also compare the power spectrum with the analytic beam model and the FEE beam model. While the effect of the beam model is negligible within the EoR window, the power spectrum with the analytic model is roughly 5\% smaller than that with the FEE model in the foreground wedge. Although it is difficult to guarantee the accuracy of the FEE model at the ultralow frequencies, the FEE model is the most reliable beam model according to recent beam modelling experiments. Thus we use the FEE model in the calibration of best data sets. 


\subsubsection{Direction Dependent Calibration}\label{sec:ddcal}

The RTS typically performs direction dependent calibration for a few of the brightest sources. In the process of DD calibration, the Jones matrix for the direction dependent calibrator is solved per each tile independently for each coarse channel. This process is intended to reduce the systematic errors from bright sources which can dominate the visibilities. However, for accurate DD calibration, a high signal to noise ratio is essential, and the number of sources to be calibrated direction dependently is a free parameter. We note that the result of the DD calibration is only used to subtract the selected DD calibrators. For the rest of sources, the result of DI calibration and ionosphere correction is applied. Further, reducing the number of DD calibrators reduces the number of degrees of freedom used in calibration and hence should only reduce any possible signal loss.

\begin{figure}
    \centering
	\includegraphics[width=8cm]{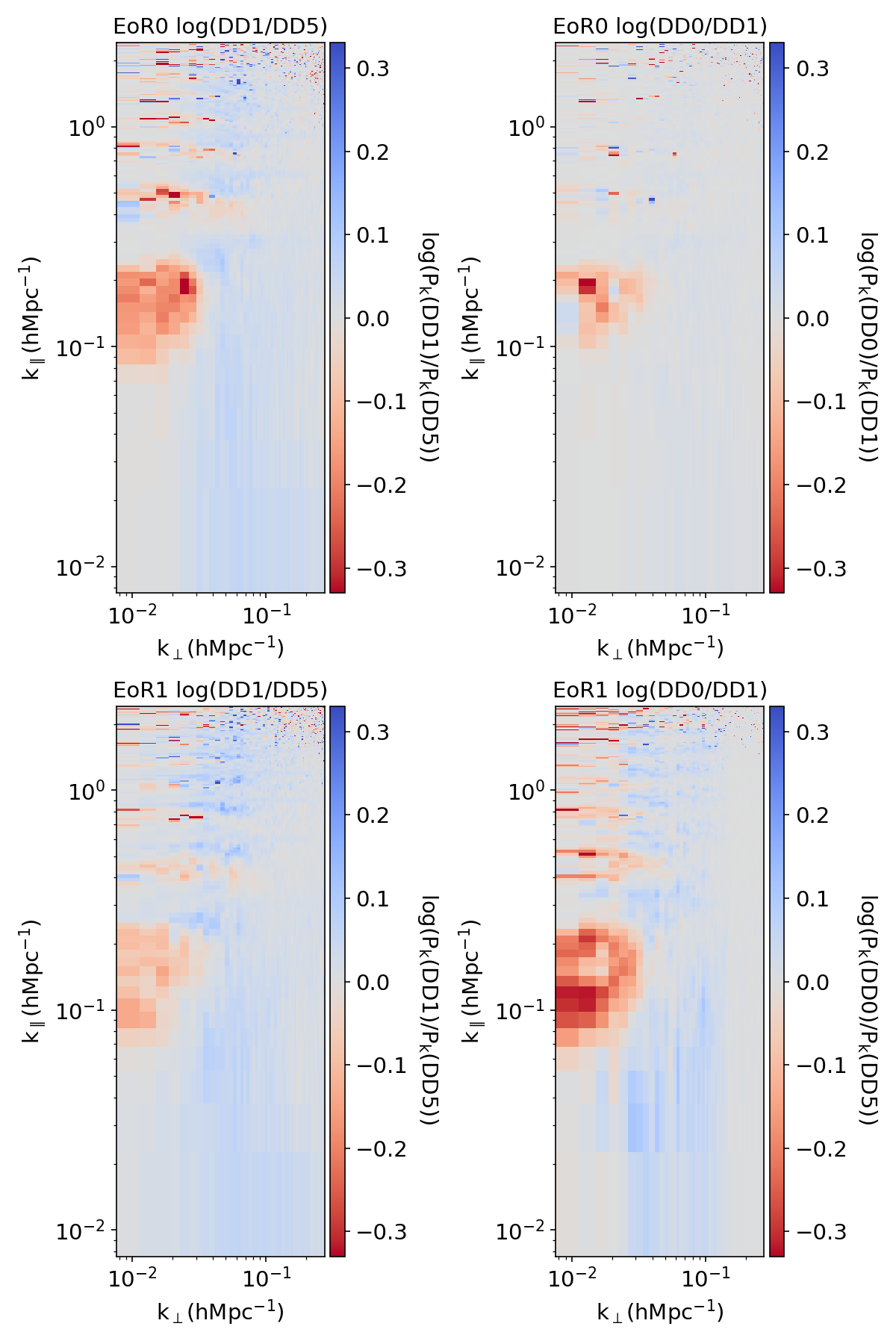}
    \caption{Comparison of power spectrum of direction dependent calibration for five sources (DD5), one source (DD1) and zero soruces (DD0). Left and right panels are the logarithm of the ratio of DD1 to DD5 and of DD0 to DD1. Top row is for EoR0 and bottom row is for EoR1. Red indicates that the power spectrum with less DD calibrators has lower power in the EoR window. }
    \label{fig:compnoDD}
\end{figure}

Using 25~min of integrated zenith observation for each the EoR0 and the EoR1 fields, we here compare the 2D power spectrum with five, one and zero DD sources. The apparent flux density of the five DD sources for the zenith observation of the EoR0 (EoR1) field are in range from 23~(34)~apparent~Jy to 35~(737)~apparent~Jy at 88~MHz. These five sources are indicated with star markers in Fig.~\ref{fig:skyimage}. We note that depending on the observation, RTS can select different DD calibrator since the beam response varies with time and pointings. In left panels of Fig.~\ref{fig:compnoDD}, we compare the 2D power spectrum with five DD sources and one DD source. An improvement is found in the EoR window between FG wedge and first coarse band harmonic. This red region means that the power spectrum with one DD source is roughly 1.5 times lower than that with 5 DD sources. This indicates that the DD calibration for the fainter 4 sources leaves additional spectral structure in the residuals. In the right panels of Fig.~\ref{fig:compnoDD}, we compare the 2D power spectrum with one DD source and without any DD sources. The power spectrum without DD calibrator shows further improvements in the EoR window especially for the EoR1 field. This indicates the spectrally non-smooth residual of Fornax~A, which is the primary DD calibrator of the EoR1, dominates these scales.  We will use zero DD calibrators in our final run since the power spectrum shows the lowest power in the EoR window.





\subsubsection{Galactic Plane}

The primary beam of the MWA is very large and the sidelobes can often produce significant response to the Galactic plane, especially at the ultralow frequencies. As the ultralow observation strategy is a drift-and-shift tracking towards the desired field, the beam response varies with pointing every 30 minutes. The different pointings are denoted by different numbers. The zenith observation is `0', and the pointings before and after zenith are `-1' and `+1' \citep{Beardsley2016FIRST7}. Here we investigate the contamination of Galactic plane and pointing dependence of the power spectrum. 

\begin{figure}
	\includegraphics[width=\columnwidth]{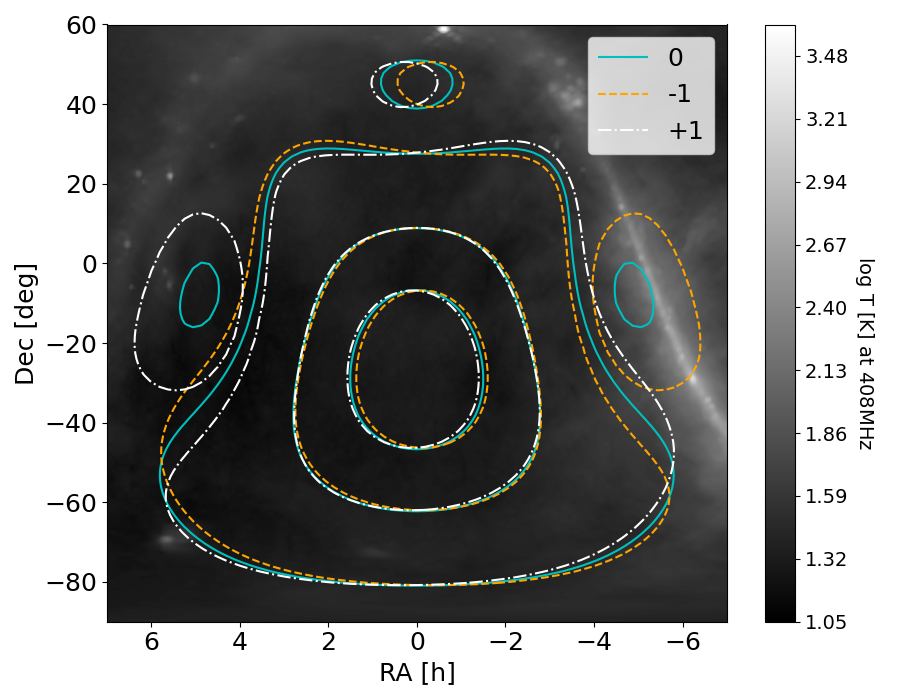}
	\includegraphics[width=\columnwidth]{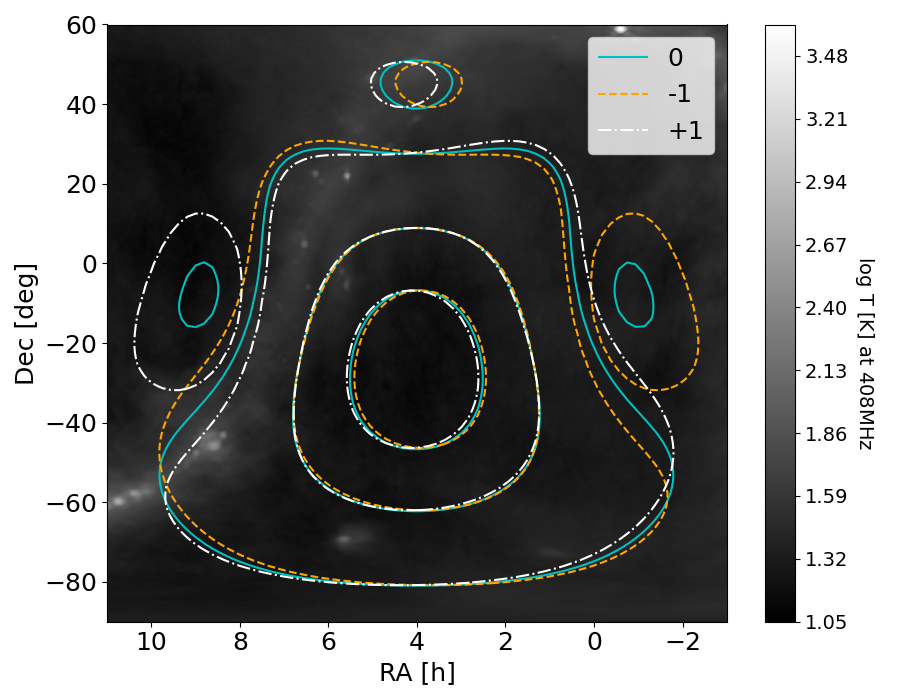}
    \caption{Beam response at 81~MHz over the 408~MHz improved version of Haslam map for EoR0 and EoR1. The contour indicates beam response of 0.001, 0.1 and 0.5. Solid line (0 pointing) is the zenith observation, the dashed (-1 pointing) and dot-dashed (+1 pointing) lines are the off-zenith observations. }
    \label{fig:BeamHaslam}
\end{figure}

 Fig.~\ref{fig:BeamHaslam} shows examples of the beam response superimposed on the Haslam map. For the EoR0 field, the first side lobe overlaps with the Galactic plane especially for the -1 pointing. For the EoR1 field, the Galactic plane at 75~MHz overlaps with the edge of the main lobe.

\begin{figure*}
    \includegraphics[width=180mm]{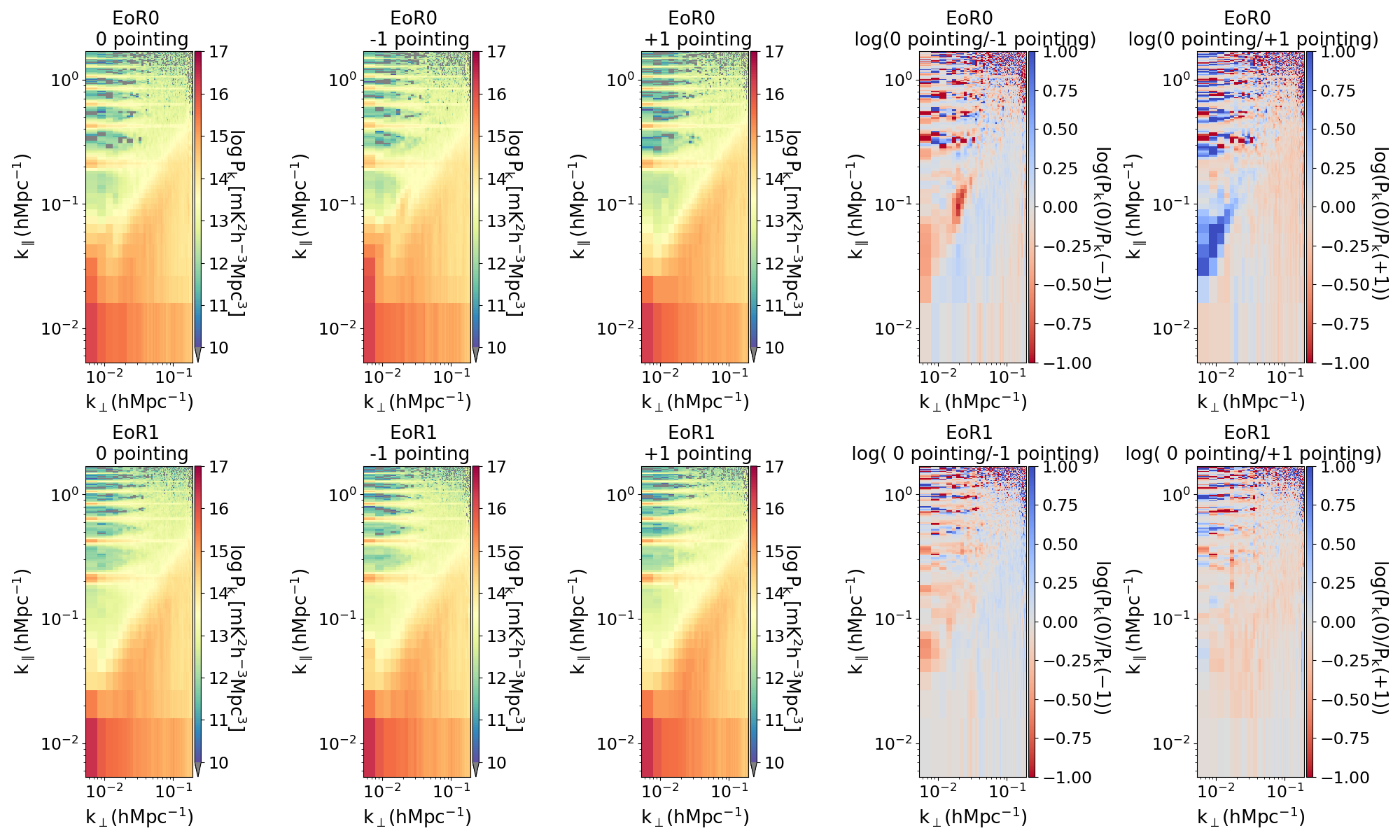}
\caption{2D power spectrum of EoR0 (top row) and EoR1 (bottom row). Compared different grid points(pointings=0, -1, +1). Each plot includes 10~min of integrated observations. These are {N-S} power spectrum. For the EoR0 of -1 pointing, strong Galactic plane contamination is found at horizon limit. The right two panels are the logarithmic ratio of 0 pointing to -1 pointing and 0 pointing to +1 pointing. Red indicates that power spectrum of 0 pointing has less power than others. }
    \label{fig:2DPSN-S}
\end{figure*}

The contamination from Galactic plane can be seen in the 2D power spectrum as shown in Fig.~\ref{fig:2DPSN-S}. We select 10~min of integrated observation with the quiet ionosphere for each pointing. 

Top row panels of Fig.~\ref{fig:2DPSN-S} show the power spectrum for the EoR0 field. Since the Galactic plane is just above the horizon, the difference between pointings are visible within the horizon limit. The side lobe propagates the Galactic plane into the FG wedge for -1 pointing. 

The power spectrum of EoR1 field is shown in bottom row panels. The difference between pointings are not as prominent as the EoR0. The Galactic plane becomes strong for -1 pointing. This is because the primary beam overlaps the Galactic anti-centre as shown in Fig.~\ref{fig:BeamHaslam}.

Note that the {N-S} polarization has powerful Galactic plane contamination compared to the {E-W} polarization. This is due to the MWA's beam responses being different for each polarization. Thus, in the Fig.~\ref{fig:2DPSN-S}, we only plot {N-S} polarization But even for the {E-W} polarization, the 2D power spectrum shows similar behaviour, but the contamination is not as powerful as for the {N-S}.

However, the ratio shows noise like behavior in the EoR window. Therefore, we do not remove any pointing from our final integration for both the EoR fields. On the other hand, the power spectrum of -1 pointing shows $\sim$10 times stronger contamination from Galactic plane compared to the zenith observation especially for the EoR0 field. In later section, we will investigate the effect of Galactic plane in our final upper limit by comparing power spectrum with -1 pointing and without it.

\subsubsection{Point source subtraction}

The source list used for patch calibration and peeling are made from a large catalogue built using PUMA \citep{Line2017PUMA:Algorithm2}. Most sources are described as a point source or multiple Gaussian sources. Some of the brightest extended sources, including Fornax~A and Pictor~A, are modelled using shapelets \citep{Line2020ModellingStudy2}. For creating the source list, the base frequency of beam model and maximum radius from the pointing centre are free parameters. Sources are ranked by the beam weighted flux density and only sources within the radius are selected. For the highband, radius of 20 degrees and central frequency of entire frequency range are chosen. However, for ultralow frequency, the optimized parameter is not known, and therefore we investigate the effect of the parameters.

\begin{figure}
    \centering
	\includegraphics[width=5cm]{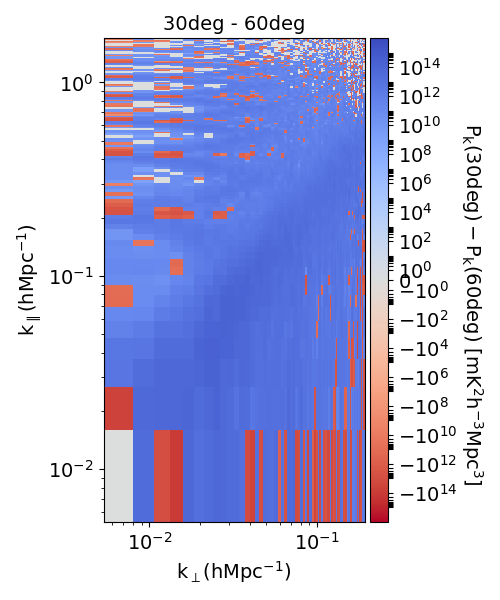}
	\caption{Diffrence plot of 2D power spectrum for EoR0 E-W with foreground subtraction within the radius of 30~deg {$P_{\rm k}(\rm 30deg)$} from the pointing centre and 60~deg {$P_{\rm k}(\rm 60deg)$}. Blue indicate $P_{\rm k}(\rm 30deg)$ has large power than $P_{\rm k}(\rm 60deg)$. }
    \label{fig:radi3060}
\end{figure}

First we demonstrate the effect of the radius. Two source lists for radius of 30 and 60 degrees are used RTS calibration separately, and the power spectrum comparison is shown in Fig.~\ref{fig:radi3060}. The difference between power spectrum with radius of 30 and 60 shows clear effect at the FG wedge. This is not surprising because a number of sources outside of the 30 degrees radius are brighter than many sources within. We also compare the source list with radius of 75 degrees, but the difference with radius of 60 degrees are negligible because these two sources lists are almost identical to each other. 

Even for the EoR0 field, sources listed in the peeling model with radius 90 degrees can pick up Pictor~A. Although Pictor~A is far from the phase center, it is bright at 80~MHz. However, even if we subtract the Pictor~A, we cannot find any improvement for 20~min integration. The subtraction of Pictor~A should be important for further calibration processing, but the source is in the edge of primary beam and the calibration has some risks due to beam correction. Thus, we do not include the Pictor~A in the peeling source list for the EoR0 data.

The base frequency has no-prominent effect compared to the radius. We make source lists with radius of 60 degrees and base frequency of 75, 80, 85 and 90~MHz. The difference of the power spectrum for each of these source list only shows noise-like difference. Thus, we use the central frequency of the full band as the base frequency of beam model.


Due to the incompleteness of the radio catalogue, no sources are listed in (RA, Dec)$\approx$(-1h, 0). At these frequencies, existing catalogues have suffered from noise and ionosphere. Thus, developing a new deep catalogue will be important for future calibration especially before the upcoming SKA1-Low observation. New surveys such as LoBES (Lynch et al in prep) and GLEAM-X (\citet[][]{2018PASA...35...33W}, Hurley-Walker et al in prep) will fill in the remaining gaps in our sky model.

\section{Power Spectrum Results}\label{results_section}

Consolidating the results of the above discussions, we attempt to diagnose a clean data set for our final power spectrum calculation. There are 446 snapshots observed in 2014. The RTS fails to generate DI calibration result for 21 observations. This might be due to some tiles not working properly. By visually inspecting the RTS DI gain phases, we omit a further 128 snapshots because they show rapid variation in gain phase spectrum.

We then further calibrate remaining data with source lists including 2000 objects. The source lists are built with a radius of 60 degrees using the beam model at 90~MHz. Regarding the source lists, we update the position of each source using the measured ionosphere offsets as in Sec.~\ref{sec:ionoupdate}. The RTS DI calibration is performed with the FEE beam model. We use zero sources for the full DD calibration (i.e. we only use the antenna gain solution obtained in the DI calibration). The point source subtraction and the ionosphere correction within the RTS are performed for 1000 sources. We list the RTS setting for the final run in Table.~\ref{tab:rtsfinal}.

\begin{table}
	\centering
	\caption{Same as the Table.~\ref{tab:rtsfid}, but for the final upper limits.}
	\label{tab:rtsfinal}
	\begin{tabular}{ll} 
		\hline
		\hline
		 Number of sources for DD calibration & 0\\ 
		 Number of sources for ionosphere correction & 1000\\ 
		 Amplitude correction & on\\
		 Number of sources to be peeled & 1000\\ 
		 Beam model & FEE model\\ 
		 Minimum baseline length & 20 $\lambda$\\ 
		 Tapered baseline length & 40 $\lambda$\\ 

		\hline
	\end{tabular}
\end{table}


In order to avoid the contamination from both the ionosphere and RFI, we separate the data into clean and polluted data sets. As the threshold, we regard the data with either the ionosphere median offset of more than 0.15~arcmin and averaged RFI occupancy of more than 3\% at higher coarse bands as the polluted data. 

\begin{figure}
    \includegraphics[width=\columnwidth]{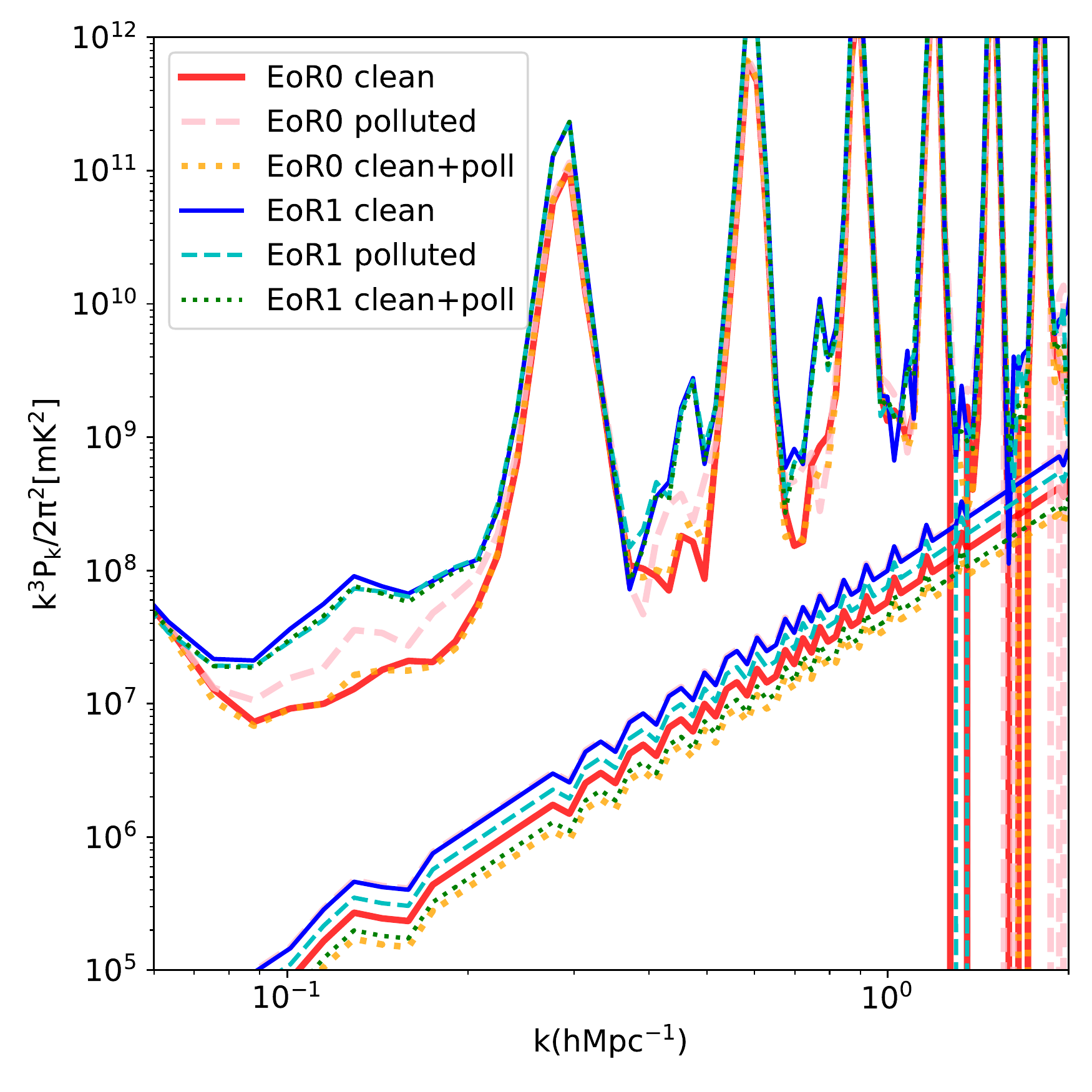}    
    \caption{Comparing the measured 1D power spectrum for the clean data, polluted data and integrating both data. We only use zenith observations and for E-W polarization. The lower straight lines are the two sigma uncertainty.  }
    \label{fig:PSclean}
\end{figure}

Fig.~\ref{fig:PSclean} shows the 1D power spectrum of the clean data, polluted data and integrating both data. For the EoR0 field, the clean data are clearly better than the polluted data at $\rm k\sim 0.1~h~Mpc ^{-1}$. This indicates the data threshold works well to improve the integration. However, the clean data has twice as many observations as the polluted data. Thus, this comparison might not be fair. But, even if we integrate the clean and polluted data, the power spectrum does not change at the scale. 

On the other hand, for the EoR1 field, the threshold does not improve the power spectrum. Rather the clean data shows slightly higher power than the polluted data, although the ratio of clean data volume to polluted data is only 0.75 in the EoR1. This seems to indicate that the power spectrum of the EoR1 field is limited by other systematics rather than the ionosphere and the RFI.

We omit the systematically polluted data from our further integration because integrating polluted data does not show improvement. Furthermore, the data with high RFI occupancy can generate unwanted bias in the power and the ionosphercally active data has biased gain value even after updating the source list. Therefore, omitting the polluted data should be worth to avoid unknown biases.  Although the RFI and ionosphere might not be dominant systematics for the EoR1 field, we remove the polluted data from further integration for a conservative analysis. 

Consequently, 2.8 hours and 2.4 hours of the best data are available for the EoR0 and the EoR1 fields, respectively. The integrated gridded visibilities are separated into three groups with respect to frequency: $\rm 75~MHz-88~MHz$, $\rm 81~MHz-94~MHz$ and $\rm 89~MHz-101~MHz$. These separated visibility boxes are used for the power spectrum estimation at each redshift. 

\begin{figure*}
    \includegraphics[width=160mm]{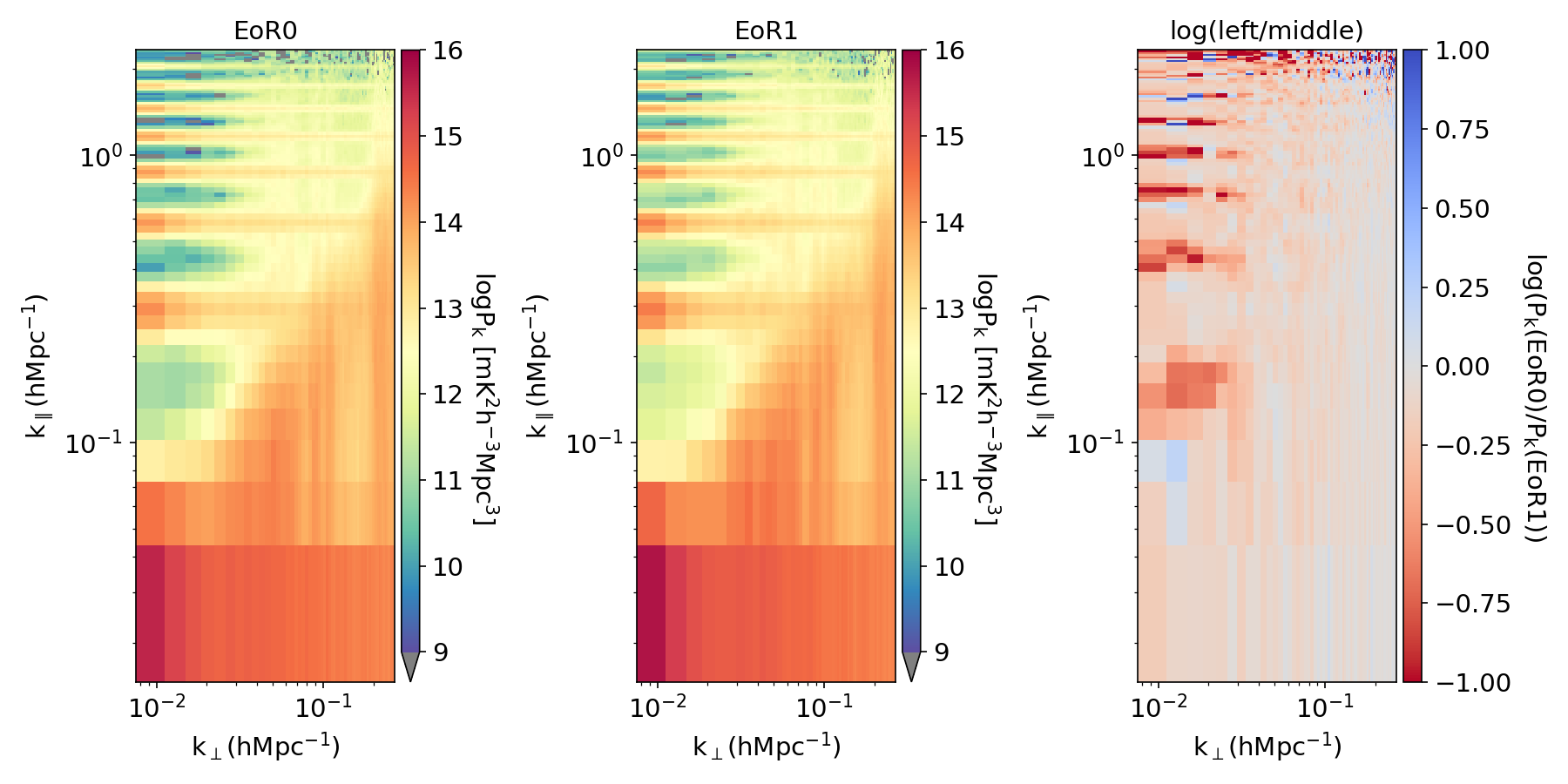}
    \caption{Measured two dimensional power spectrum of the best clean data set for E-W polarization for the EoR0 (left) and the EoR1 (right). The frequency range is 75~MHz to 87~MHz. Right panel is the logarithm of the ratio between the EoR0 and the EoR1. Red indicates that the power spectrum for the EoR0 is lower than the power spectrum for the EoR1.}
    \label{fig:2DPSfield}
\end{figure*}

Fig.~\ref{fig:2DPSfield} shows the 2D power spectrum of the best data sets for the E-W polarization. The EoR0 field is better than the EoR1 field in the EoR window. Since the difference is prominent at higher $\rm k_{\parallel}$, this indicates there is additional spectrally non-smooth contamination for the EoR1 field. This may be due to contamination from residuals of subtracted bright sources, as has been found for the EoR1 field at higher frequencies (Rahimi et al., in preparation). 

\begin{figure*}
    \includegraphics[width=170mm]{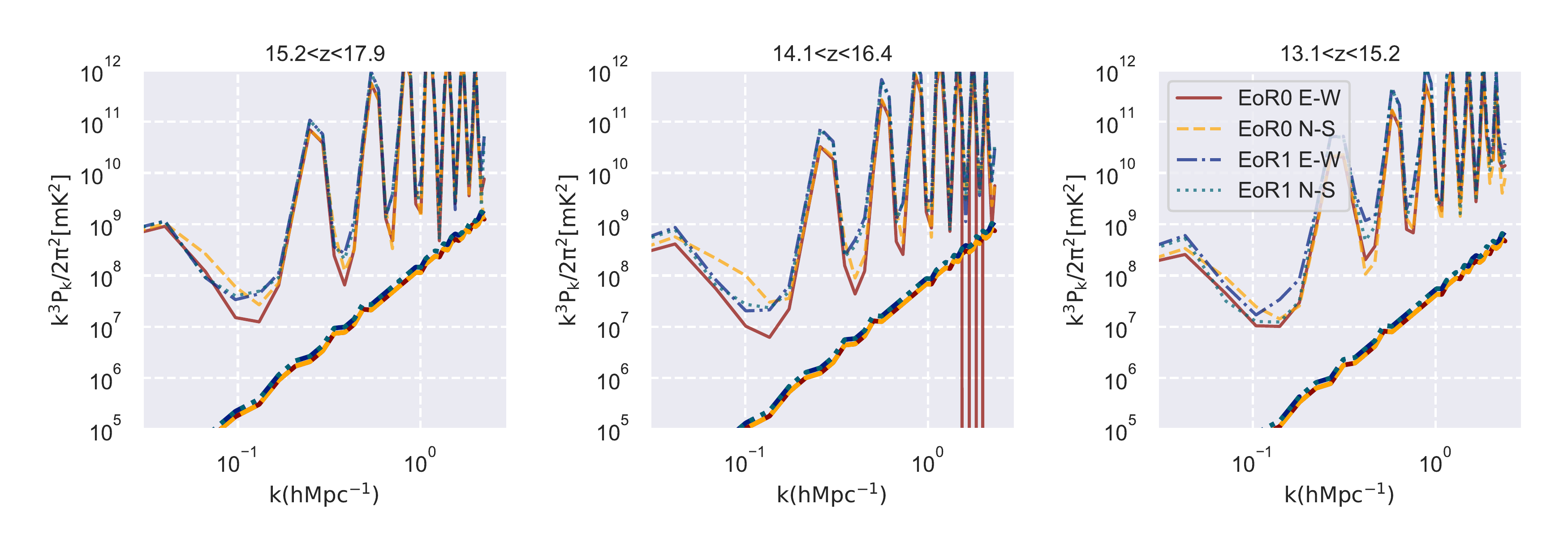}
    \caption{Measured power spectrum of the best clean data sets for each redshift. The solid, dashed, dot-dashed and dotted lines are the EoR0 for E-W polarization, EoR0 N-S, EoR1 E-W and EoR1 N-S respectively. Thick lines show two sigma thermal uncertainties.}
    \label{fig:1Dupper}
\end{figure*}

Fig.~\ref{fig:1Dupper} shows the measured 1D power spectrum for each frequency group. The EoR0 for E-W polarization has the lowest power spectrum at $\rm k=0.1~h~Mpc^{-1}$. Due to the galactic plane, the power spectrum of EoR0 for N-S polarization is larger than the E-W polarization. Note that the power spectrum is reduced by a factor of 1.15 at $\rm k<0.07 \rm ~h~Mpc^{-1}$ if we omit the data of -1 pointing for the EoR0 field. The power spectrum of EoR1 is roughly 3 times larger than the EoR0 at $\rm k\sim 0.1 \rm ~h~Mpc^{-1}$ and at $z=16.5$. In Table~\ref{tab:upper}, we summarise the best 2 sigma upper limits for each field and polarization.

\begin{table*}
	\centering
	\caption{Two sigma upper limits on the amplitude of the 21~cm signal in units of $\rm mK^2$ for each EoR field, polarization and redshift. The result of DI calibration is used, and the RTS setting for the limit is listed in Table~\ref{tab:rtsfinal}. There is a possibility of a few \% of signal loss due to DI calibration for the EoR0 field.}
	\label{tab:upper}
	\begin{tabular}{ccccccc} 
		\hline
		k[$\rm ~h~Mpc^{-1}$] &redshift & & EoR0 E-W & EoR0 N-S & EoR1 E-W & EoR1 N-S \\
		\hline
		9.7e-02 & $z$=16.5 &(15.2<z<17.9)& 1.5e+07 & 6.0e+07 & 3.4e+07 &4.0e+07   \\ 
		1.3e-01 & $z$=16.5&(15.2<z<17.9)& 1.3e+07 & 2.7e+07 & 4.4e+07 &5.0e+07   \\ 
		\hline
		1.0e-01 & $z$=15.2&(14.1<z<16.4)& 1.0e+07 & 9.9e+07 & 2.0e+07 &2.7e+07  \\
        1.4e-01 & $z$=15.2&(14.1<z<16.4)& 6.3e+06 & 3.0e+07 & 2.1e+07 &2.4e+07  \\
        \hline
		1.0e-01 & $z$=14.2&(13.1<z<15.2)& 1.0e+07 & 2.5e+07 & 1.7e+07 &1.3e+07  \\
        1.4e-01 & $z$=14.2&(13.1<z<15.2)& 1.0e+07 & 1.4e+07 & 3.4e+07 &1.2e+07 \\
		\hline
	\end{tabular}
\end{table*}

We finally address the possibility of signal loss. DD calibration can lead significant signal loss at short baselines if Galactic diffuse emission is not included in the calibration model \citep[e.g.][]{2016MNRAS.463.4317P}. One way to mitigate the signal loss is the exclusion of short baselines from the calibration although this cut can cause an enhancement in noise. We emphasize that the signal loss due to DD calibration should not affect our final upper limits because the DD Jones matrix calibration is not used. Furthermore, as shown in Sec~\ref{sec:ddcal}, the upper limits are enhanced as increasing the number of DD calibration sources. Thus, the signal loss associated with DD calibration is trivial in our final limits. On the other hand, however, the result of DI calibration applied to our final result can depend on the threshold wavelength as short baselines less than 20 $\lambda$ and 40 $\lambda$ are excluded and tapered in the DI calibration. To investigate this issue, we compare the angular power spectrum by varying the threshold value using 10 snapshots. Following \citep[e.g.][]{2016MNRAS.463.4317P}, we exclude baselines shorter than 200 $\lambda$, and then the angular power spectrum at $|{\bf{u}}|<200$ shows an enhancement of less than 10\% for the EoR0 field. Further signal loss may not exist because the value does not change for higher threshold values. For the EoR1 field, on the other hand, we find roughly 5\% reduction in the angular power spectrum with high threshold wavelength at $u<50$. The reduction might indicate a bias due to the error of the calibration catalogue because significantly bright sources dominate the diffuse emission even at $u=10$ in the EoR1 field. It should be mentioned that because the sensitivity decreases when increasing the minimum baseline length, the antenna gain results have larger errors and become more susceptible to the systematics on long baselines, and hence it is hard to tell if the bias is due to signal loss or other systematic errors. For more concrete  analysis, an end-to-end simulation including diffuse emission is required to probe it. Building such a large simulation is beyond the scope of this paper. Thus, we only caution there are a few \% of signal loss on our upper limits. 


\section{Discussion and Conclusion}\label{sec:discussion}

By analysing 15 hours of MWA Phase~I ultralow data, we have  investigated and mitigated against several systematic errors that dominate at these lower frequencies. Our findings can be summarised as follows: (i) While only 0.5\% of the lower bands (75~MHz-85~MHz) are contaminated by the RFI, the FM radio occupies more than 10\% of data at many fine channels. (ii) Since the ionospheric effect becomes stronger with $\nu^{-2}$, the ionosphere makes modelling errors in the direction independent calibration. Updating the calibration model reduces the systematics, but further implementation will be required to remove the error completely. (iii) Due to the wide field of view, a calibration model has to include radio sources within a radius of 60 deg from pointing centre. (iv) Because the primary beam side lobe overlaps the Galactic plane for the EoR0, the power spectrum of N-S polarization shows prominent Galactic plane contamination at the horizon limit of the foreground wedge. (v) The direction dependent calibration leaves spectrally non-smooth residuals after the point source subtraction as shown in Sec.~\ref{sec:ddcal} and therefore we used zero direction dependent calibrators and the residual contamination is mitigated in the EoR window.

Based on the above jackknife analysis, we optimize the calibration of the RTS and remove systematic polluted data using thresholds on the RFI occupancy and ionosphere condition. We could use roughly 35\% of all data as clean data sets. By calculating the two dimensional power spectrum using the clean data, we find the EoR window of EoR1 is highly contaminated compared to the EoR0. The source of difference is not revealed but the two brightest sources (Fornax~A and Pictor~A) might contaminate the spectral fluctuation. For the spherical averaged 1 dimensional power spectrum, the systematics dominated best upper limits of roughly $10^7 ~ \rm mK^2$ is obtained for the EoR0 and for the E-W polarization at $13< z < 18$ with a possibility of signal loss of a few \%. 


We mention here that the evolution of 21~cm signal along line of sight within the band width which is called the light-cone effect. Ignoring this effect can bias the cosmological power spectrum by a factor of 2 at the cosmic dawn ($z\approx$18) with the redshift range $\Delta z$ of 2.5 \citep{2015MNRAS.453.3143G,2018MNRAS.477.3217G}. The effect is important at timing of a rapid evolution which may have happened at around $z=18$ as indicated by the strong absorption in \cite{Bowman2018AnSpectrum}. Thus, power spectrum analysies at these low frequencies need to address this effect.

In Sec.~\ref{results_section}, we calculated the power spectrum using gridded visibility data with bandwidth of 12.8~MHz. This frequency range is wider than that in \cite{Eastwood2019TheOVRO-LWA}, they used only 2.6~MHz. Since we apply Blackman-Nuttall window \citep[e.g. equation (17) in][]{2013ApJ...776....6T}, the effective bandwidth becomes roughly 4.65~MHz.  Due to down weighting at the edges of the 12.8~MHz band, the effective redshift range is narrowed as well. While the exact redshift range is not clear due to the taper, it should be $\Delta z \approx 1.0$ at $z=16.6$. 

\begin{figure}
    \includegraphics[width=80mm]{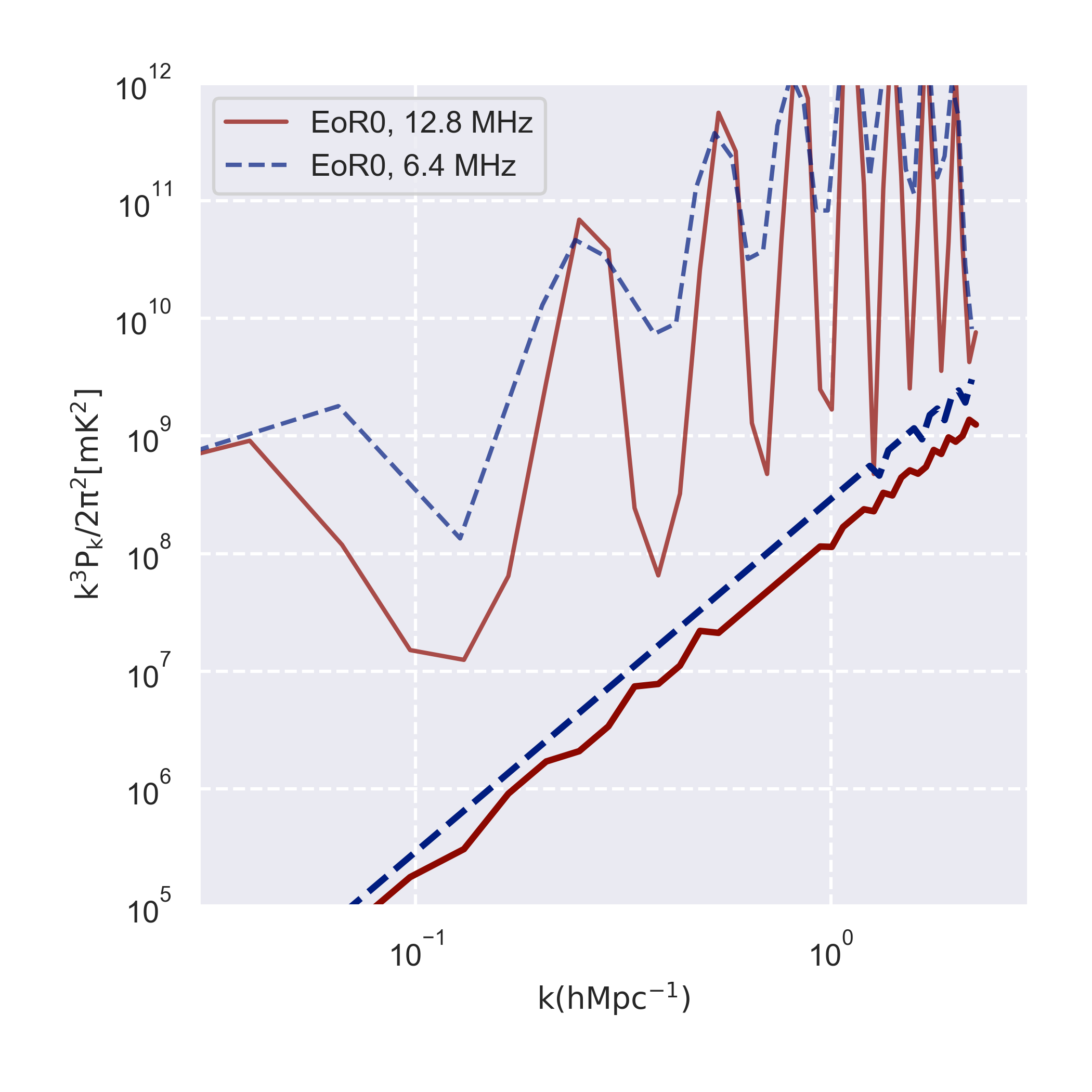}
    \caption{ Measured power spectrum of the best clean data sets at $z=16.5$ for the EoR0 field. The solid and dashed lines are the results with bandwidths of 12.8~MHz and 6.4~MHz respectively. Thick lines show two sigma thermal uncertainties.}
    \label{fig:1Duppernarros}
\end{figure}

As an experiment, we calculated power spectrum using gridded visibility with frequency range from 78.1~MHz to 84.5~MHz. The bandwidth of 6.4~MHz should correspond to $\Delta z \approx 0.5$ at $z=16.6$. Fig.~\ref{fig:1Duppernarros} shows that  the power spectrum with bandwidth of 6.4~MHz is 10 times larger than 12.8~MHz. Since band width becomes half, the resolution in $\rm k_{\parallel}$ becomes worse, and the coarse band channel harmonics cannot be well separated each other. Since the enhancement seems more drastic compared to the light-cone effect, we provided the upper limits with wider bandwidth. 

The direction dependent (DD) effect on the direction to sources omitted from DD calibrator should be solved. When we peel such sources, ionospheric effect is corrected and the amplitude of model is just scaled by a factor over the all coarse bands before the source subtraction. Therefore the DD calibration using a smooth polynomial function over the all coarse bands might be worth. For example, such analysis is done in \cite{2020MNRAS.493.1662M}. They solved DD gain calibration with a third order Bernstein polynomial so that the solution is spectrally smooth. They expected that non-smooth instrumental response is well corrected by the result of DI calibration and also assume that DD effects (e.g. ionosphere and beam response) are spectrallly smooth. These assumption should also be valid in our case. Developing new implementation of the RTS is beyond the purpose of this paper. Thus, we left such modification for future work.

Instrumental systematics and exotic physical models regarding the Cosmic Dawn have been suggested to explain the EDGES absorption feature. Our upper limit is still $\sim 1-2$ orders of magnitude too high to strongly constrain such models. However, the upper limit obtained in this work is an order of magnitude lower than the previous analysis EW16. This improvement encourages further processing of ultralow data. Upcoming MWA Phase~III observation performed with 256 tiles simultaneously, and the MWA Phase~III has higher sensitivity and higher resolution compared to the MWA Phase~I. Thus the future observation at ultralow frequencies would reduce the upper limits further by providing improved radio catalogues and reducing instrumental systematics, and the result will give us important insights on the EDGES results and the Cosmic Dawn. Furthermore, as we could find systematics such as the RFI condition, gain error due to ionosphere and wide field of view, continuing the data analysis at the lowest band would reveal unknown systematics and be worthwhile for future analysis of SKA1-Low data. 


\section*{Data Availabilty}

The data used in this work are publicly available from the MWA archive via \url{https://asvo.mwatelescope.org/}

\section*{Acknowledgements}


We would like to thank to anonymous referee for his/her useful comments. We also thank to Dr. Padmanabhan for her comments on early version of this paper. This research was supported by the Australian Research Council Centre of Excellence for All Sky Astrophysics in 3 Dimensions (ASTRO 3D), through project number CE170100013. SY is supported by JSPS Overseas Research Fellowships and this work was supported in part by JSPS KAKENHI. Grants: 16H05999 (SY). CMT is supported by an ARC Future Fellowship through project number FT180100321. Computational resources used were awarded under Astronomy Australia Ltd’s ASTAC merit allocation scheme on the OzSTAR national facility at Swinburne University of Technology. The OzSTAR program receives funding in part from the Astronomy National Collaborative Research Infrastructure Strategy (NCRIS) allocation provided by the Australian Government. We acknowledge the Pawsey Supercomputing Centre which is supported by the Western Australian and Australian Governments. This scientific work makes use of the Murchison Radio-astronomy Observatory, operated by CSIRO. We acknowledge the Wajarri Yamatji people as the traditional owners of the Observatory site. 
We acknowledge the use of the Legacy Archive for Microwave Background Data Analysis (LAMBDA), part of the High Energy Astrophysics Science Archive Center (HEASARC). HEASARC/LAMBDA is a service of the Astrophysics Science Division at the NASA Goddard Space Flight Center.



\bibliographystyle{mnras}
\bibliography{eor,eorSY} 




\appendix

\section{Calculated Power Spectrum And Thermal Noise Uncertainty}

We here summarize all power spectrum at ultralow data for comparing in future analysis. In table~\ref{tab:app1}, \ref{tab:app2} and \ref{tab:app3}, we show the results at $z=16.5$, 15.2 and 14.2.  The list contains the measured power spectrum and 2$\sigma$ thermal noise for each poralization and each EoR field. As presented in Sec~\ref{results_section}, the results are estimated from 2.8 hours and 2.4 hours of the best clean data. 

\begin{table*}
	\centering
	\caption{The measured power spectrum and 2$\sigma$ thermal noise uncertainty in units of $\rm mK^2$ for each EoR field, polarization at $z=16.5$. Same as Table.~\ref{tab:upper}, the result of DI calibration is used, and the RTS setting for the result is Table~\ref{tab:rtsfinal}. There is a possibility of signal loss of a few \% due to DI calibration.}
	\label{tab:app1}
	\begin{tabular}{ccccccccc} 
		\hline
		\multicolumn{1}{|c|}{$z=16.5$ (15.2<$z$<17.9)} & \multicolumn{4}{|c|}{EoR0} & \multicolumn{4}{|c|}{EoR1} \\
		\hline
		k $\rm ~h~Mpc^{-1}$ & $\Delta^2_{\rm N-S}$ &  $\Delta^2_{\rm E-W}$ & $\rm Thermal_{\rm N-S}$ & $\rm Thermal_{\rm E-W}$ &  $\Delta^2_{\rm N-S}$  &  $\Delta^2_{\rm E-W}$  &  $\rm Thermal_{\rm N-S}$ & $\rm Thermal_{\rm E-W}$ \\
		\hline
\hline
1.8e-02 & 4.5e+08 & 4.5e+08 & 6.9e+02 & 6.9e+02 & 5.7e+08 & 5.7e+08 & 8.7e+02 &8.7e+02 \\ 
4.0e-02 & 9.1e+08 & 1.0e+09 & 4.7e+03 & 4.7e+03 & 1.2e+09 & 1.1e+09 & 5.9e+03 &5.9e+03 \\ 
6.7e-02 & 1.2e+08 & 2.6e+08 & 2.2e+04 & 2.2e+04 & 9.1e+07 & 9.1e+07 & 2.7e+04 &2.7e+04 \\ 
9.7e-02 & 1.5e+07 & 6.0e+07 & 6.7e+04 & 6.7e+04 & 3.4e+07 & 4.0e+07 & 8.5e+04 &8.5e+04 \\ 
1.3e-01 & 1.2e+07 & 2.7e+07 & 1.2e+05 & 1.2e+05 & 4.4e+07 & 4.9e+07 & 1.5e+05 &1.5e+05 \\ 
1.7e-01 & 6.4e+07 & 7.2e+07 & 3.4e+05 & 3.4e+05 & 1.1e+08 & 9.2e+07 & 4.4e+05 &4.4e+05 \\ 
2.1e-01 & 2.9e+09 & 2.7e+09 & 6.5e+05 & 6.5e+05 & 4.3e+09 & 4.1e+09 & 8.2e+05 &8.2e+05 \\ 
2.5e-01 & 6.9e+10 & 7.0e+10 & 7.9e+05 & 7.9e+05 & 1.1e+11 & 1.0e+11 & 1.0e+06 &1.0e+06 \\ 
2.9e-01 & 3.8e+10 & 4.0e+10 & 1.3e+06 & 1.3e+06 & 5.7e+10 & 5.6e+10 & 1.6e+06 &1.6e+06 \\ 
3.4e-01 & 2.4e+08 & 7.2e+08 & 2.8e+06 & 2.8e+06 & 3.6e+08 & 4.0e+08 & 3.6e+06 &3.6e+06 \\ 
3.8e-01 & 6.5e+07 & 1.2e+08 & 2.9e+06 & 2.9e+06 & 2.7e+08 & 2.1e+08 & 3.7e+06 &3.7e+06 \\ 
4.3e-01 & 3.2e+08 & 3.1e+08 & 4.2e+06 & 4.2e+06 & 1.2e+09 & 8.5e+08 & 5.4e+06 &5.4e+06 \\ 
4.8e-01 & 2.6e+10 & 2.6e+10 & 8.3e+06 & 8.3e+06 & 4.7e+10 & 4.4e+10 & 1.1e+07 &1.1e+07 \\ 
5.4e-01 & 5.6e+11 & 5.6e+11 & 8.0e+06 & 8.0e+06 & 9.5e+11 & 9.1e+11 & 1.0e+07 &1.0e+07 \\ 
5.9e-01 & 2.6e+11 & 2.7e+11 & 1.1e+07 & 1.1e+07 & 4.1e+11 & 3.9e+11 & 1.4e+07 &1.4e+07 \\ 
6.5e-01 & 1.3e+09 & 2.0e+09 & 1.4e+07 & 1.4e+07 & 1.4e+09 & 1.4e+09 & 1.8e+07 &1.8e+07 \\ 
7.0e-01 & 4.7e+08 & 3.3e+08 & 1.8e+07 & 1.8e+07 & 3.8e+09 & 3.7e+09 & 2.3e+07 &2.3e+07 \\ 
7.6e-01 & 4.6e+10 & 4.4e+10 & 2.3e+07 & 2.3e+07 & 7.7e+10 & 7.2e+10 & 2.9e+07 &2.9e+07 \\ 
8.2e-01 & 1.7e+12 & 1.7e+12 & 2.9e+07 & 2.9e+07 & 2.7e+12 & 2.6e+12 & 3.6e+07 &3.6e+07 \\ 
8.8e-01 & 7.4e+11 & 7.7e+11 & 3.6e+07 & 3.6e+07 & 1.1e+12 & 1.1e+12 & 4.5e+07 &4.5e+07 \\ 
9.4e-01 & 2.5e+09 & 5.6e+09 & 4.4e+07 & 4.4e+07 & 3.3e+09 & 3.3e+09 & 5.5e+07 &5.5e+07 \\ 
		\hline
	\end{tabular}
\end{table*}

\begin{table*}
	\centering
	\caption{Same as \ref{tab:app1}, but for $z=15.2$.}
	\label{tab:app2}
	\begin{tabular}{ccccccccc} 
		\hline
		\multicolumn{1}{|c|}{$z=15.2$ (14.1<$z$<16.4)} & \multicolumn{4}{|c|}{EoR0} & \multicolumn{4}{|c|}{EoR1} \\
		\hline
		k $\rm ~h~Mpc^{-1}$ & $\Delta^2_{\rm N-S}$ &  $\Delta^2_{\rm E-W}$ & $\rm Thermal_{\rm N-S}$ & $\rm Thermal_{\rm E-W}$ &  $\Delta^2_{\rm N-S}$  &  $\Delta^2_{\rm E-W}$  &  $\rm Thermal_{\rm N-S}$ & $\rm Thermal_{\rm E-W}$ \\
		\hline
\hline
1.9e-02 & 1.9e+08 & 2.1e+08 & 3.9e+02 & 3.9e+02 & 3.1e+08 & 3.1e+08 & 4.9e+02 &4.9e+02  \\
4.1e-02 & 4.1e+08 & 5.7e+08 & 2.7e+03 & 2.7e+03 & 8.7e+08 & 7.8e+08 & 3.5e+03 &3.5e+03  \\
6.9e-02 & 5.5e+07 & 2.2e+08 & 1.3e+04 & 1.3e+04 & 8.2e+07 & 6.4e+07 & 1.6e+04 &1.6e+04  \\
1.0e-01 & 1.0e+07 & 9.9e+07 & 4.0e+04 & 4.0e+04 & 2.0e+07 & 2.7e+07 & 5.1e+04 &5.1e+04  \\
1.4e-01 & 6.2e+06 & 3.0e+07 & 6.9e+04 & 6.9e+04 & 2.1e+07 & 2.3e+07 & 8.8e+04 &8.8e+04  \\
1.7e-01 & 2.2e+07 & 3.6e+07 & 2.0e+05 & 2.0e+05 & 6.3e+07 & 4.8e+07 & 2.6e+05 &2.6e+05  \\
2.1e-01 & 1.2e+09 & 1.2e+09 & 3.8e+05 & 3.8e+05 & 2.7e+09 & 2.4e+09 & 4.9e+05 &4.9e+05  \\
2.6e-01 & 3.3e+10 & 3.4e+10 & 4.7e+05 & 4.7e+05 & 7.4e+10 & 7.0e+10 & 6.0e+05 &6.0e+05  \\
3.0e-01 & 1.8e+10 & 2.1e+10 & 7.6e+05 & 7.6e+05 & 4.1e+10 & 3.9e+10 & 9.7e+05 &9.7e+05  \\
3.5e-01 & 1.6e+08 & 4.7e+08 & 1.7e+06 & 1.7e+06 & 2.4e+08 & 2.3e+08 & 2.1e+06 &2.1e+06  \\
4.0e-01 & 4.4e+07 & 8.9e+07 & 1.8e+06 & 1.8e+06 & 5.0e+08 & 3.7e+08 & 2.2e+06 &2.2e+06  \\
4.5e-01 & 1.2e+08 & 2.4e+08 & 2.5e+06 & 2.5e+06 & 1.3e+09 & 1.0e+09 & 3.2e+06 &3.2e+06  \\
5.0e-01 & 1.1e+10 & 1.2e+10 & 5.0e+06 & 5.0e+06 & 3.0e+10 & 2.5e+10 & 6.3e+06 &6.3e+06  \\
5.6e-01 & 2.6e+11 & 2.7e+11 & 4.8e+06 & 4.8e+06 & 6.6e+11 & 6.1e+11 & 6.1e+06 &6.1e+06  \\
6.1e-01 & 1.2e+11 & 1.4e+11 & 6.4e+06 & 6.4e+06 & 3.1e+11 & 2.8e+11 & 8.1e+06 &8.1e+06  \\
6.7e-01 & 4.6e+08 & 1.9e+09 & 8.3e+06 & 8.3e+06 & 1.3e+09 & 1.1e+09 & 1.1e+07 &1.1e+07  \\
7.3e-01 & 2.6e+08 & 4.3e+08 & 1.1e+07 & 1.1e+07 & 2.6e+09 & 2.8e+09 & 1.4e+07 &1.4e+07  \\
7.9e-01 & 2.0e+10 & 2.0e+10 & 1.4e+07 & 1.4e+07 & 5.2e+10 & 4.3e+10 & 1.7e+07 &1.7e+07  \\
8.5e-01 & 8.1e+11 & 8.4e+11 & 1.7e+07 & 1.7e+07 & 1.9e+12 & 1.8e+12 & 2.2e+07 &2.2e+07  \\
9.1e-01 & 3.6e+11 & 4.0e+11 & 2.1e+07 & 2.1e+07 & 8.0e+11 & 7.4e+11 & 2.7e+07 &2.7e+07  \\
9.8e-01 & 1.7e+09 & 5.3e+09 & 2.6e+07 & 2.6e+07 & 2.9e+09 & 1.8e+09 & 3.3e+07 &3.3e+07  \\
		\hline
	\end{tabular}
\end{table*}

\begin{table*}
	\centering
	\caption{Same as Table~\ref{tab:app3}, but for $z=14.2$.}
	\label{tab:app3}
	\begin{tabular}{ccccccccc} 
		\hline
		\multicolumn{1}{|c|}{$z=14.2$  (13.1<$z$<15.2)} & \multicolumn{4}{|c|}{EoR0} & \multicolumn{4}{|c|}{EoR1} \\
		\hline
		k $\rm ~h~Mpc^{-1}$ & $\Delta^2_{\rm N-S}$ &  $\Delta^2_{\rm E-W}$ & $\rm Thermal_{\rm N-S}$ & $\rm Thermal_{\rm E-W}$ &  $\Delta^2_{\rm N-S}$  &  $\Delta^2_{\rm E-W}$  &  $\rm Thermal_{\rm N-S}$ & $\rm Thermal_{\rm E-W}$ \\
		\hline
\hline
1.9e-02 & 1.4e+08 & 1.4e+08 & 2.3e+02 & 2.3e+02 & 2.3e+08 & 2.1e+08 & 3.0e+02 &3.0e+02 \\
4.2e-02 & 2.6e+08 & 3.4e+08 & 1.7e+03 & 1.7e+03 & 6.0e+08 & 5.3e+08 & 2.2e+03 &2.2e+03 \\
7.1e-02 & 4.4e+07 & 9.0e+07 & 8.0e+03 & 8.0e+03 & 6.3e+07 & 3.2e+07 & 1.0e+04 &1.0e+04 \\
1.0e-01 & 1.0e+07 & 2.5e+07 & 2.5e+04 & 2.5e+04 & 1.7e+07 & 1.3e+07 & 3.2e+04 &3.2e+04 \\
1.4e-01 & 1.0e+07 & 1.4e+07 & 4.3e+04 & 4.3e+04 & 3.4e+07 & 1.2e+07 & 5.6e+04 &5.6e+04 \\
1.8e-01 & 2.9e+07 & 2.6e+07 & 1.3e+05 & 1.3e+05 & 8.5e+07 & 3.2e+07 & 1.6e+05 &1.6e+05 \\
2.2e-01 & 8.7e+08 & 7.3e+08 & 2.4e+05 & 2.4e+05 & 1.9e+09 & 1.7e+09 & 3.1e+05 &3.1e+05 \\
2.7e-01 & 2.0e+10 & 2.1e+10 & 3.0e+05 & 3.0e+05 & 5.1e+10 & 5.0e+10 & 3.8e+05 &3.8e+05 \\
3.1e-01 & 2.1e+10 & 2.3e+10 & 6.8e+05 & 6.8e+05 & 5.2e+10 & 5.0e+10 & 8.7e+05 &8.7e+05 \\
3.6e-01 & 1.7e+09 & 2.2e+09 & 7.4e+05 & 7.4e+05 & 4.1e+09 & 3.3e+09 & 9.5e+05 &9.5e+05 \\
4.1e-01 & 2.0e+08 & 1.1e+08 & 1.1e+06 & 1.1e+06 & 1.2e+09 & 4.8e+08 & 1.4e+06 &1.4e+06 \\
4.7e-01 & 3.7e+08 & 1.8e+08 & 1.6e+06 & 1.6e+06 & 1.4e+09 & 9.0e+08 & 2.0e+06 &2.0e+06 \\
5.2e-01 & 7.7e+09 & 6.7e+09 & 3.1e+06 & 3.1e+06 & 2.0e+10 & 1.9e+10 & 4.0e+06 &4.0e+06 \\
5.8e-01 & 1.6e+11 & 1.7e+11 & 3.0e+06 & 3.0e+06 & 4.5e+11 & 4.4e+11 & 3.9e+06 &3.9e+06 \\
6.3e-01 & 7.7e+10 & 8.7e+10 & 4.0e+06 & 4.0e+06 & 2.1e+11 & 2.0e+11 & 5.1e+06 &5.1e+06 \\
6.9e-01 & 7.9e+08 & 1.3e+09 & 5.2e+06 & 5.2e+06 & 2.0e+09 & 1.3e+09 & 6.7e+06 &6.7e+06 \\
7.6e-01 & 6.8e+08 & 7.9e+08 & 6.7e+06 & 6.7e+06 & 5.2e+09 & 4.1e+09 & 8.7e+06 &8.7e+06 \\
8.2e-01 & 1.4e+10 & 1.2e+10 & 8.6e+06 & 8.6e+06 & 3.5e+10 & 3.0e+10 & 1.1e+07 &1.1e+07 \\
8.8e-01 & 5.2e+11 & 5.3e+11 & 1.1e+07 & 1.1e+07 & 1.3e+12 & 1.3e+12 & 1.4e+07 &1.4e+07 \\
9.5e-01 & 2.3e+11 & 2.5e+11 & 1.3e+07 & 1.3e+07 & 5.5e+11 & 5.4e+11 & 1.7e+07 &1.7e+07 \\
1.0e+00 & 4.1e+09 & 3.1e+09 & 1.6e+07 & 1.6e+07 & 3.6e+09 & 2.8e+09 & 2.1e+07 &2.1e+07 \\
		\hline
	\end{tabular}
\end{table*}

\bsp	
\label{lastpage}
\end{document}